# cardiGAN: A Generative Adversarial Network Model for Design and Discovery of Multi Principal Element Alloys

Z. Li[1], W.T. Nash[2], S.P. O'Brien[3], Y. Qiu[2], R.K. Gupta[3], N. Birbilis[1,*]

[1]College of Engineering and Computer Science, The Australian National University, Acton, A.C.T., 2601, Australia.

[2]Department of Materials Science and Engineering, Monash University, Clayton, Victoria, 3800, Australia

[3]Department of Materials Science and Engineering, North Carolina State Unversity, Raleigh, NC, 27695, USA.

*nick.birbilis@anu.edu.au

**Abstract**

Multi-principal element alloys (MPEAs), inclusive of high entropy alloys (HEAs), continue to attract significant research attention owing to their potentially desirable properties. Although MPEAs remain under extensive research, traditional (i.e. empirical) alloy production and testing is both costly and time-consuming, partly due to the inefficiency of the early discovery process which involves experiments on a large number of alloy compositions. It is intuitive to apply machine learning in the discovery of this novel class of materials, of which only a small number of potential alloys has been probed to date. In this work, a proof-of-concept is proposed, combining generative adversarial networks (GANs) with discriminative neural networks (NNs), to accelerate the exploration of novel MPEAs. By applying the GAN model herein, it was possible to directly generate novel compositions for MPEAs, and to predict their phases. To verify the predictability of the model, alloys designed by the model are presented and a candidate produced – as validation. This suggests that the model herein offers an approach that can significantly enhance the capacity and efficiency of development of novel MPEAs.

**Keywords**: alloy design, machine learning, GAN, neural network, MPEA, high entropy alloys.

# 1. Introduction

The so-called high entropy alloys (HEAs) have gained significant popularity in the context of contemporary metallurgy research [1-12]. The simplified description of HEAs can be given as alloys comprised of, notionally, five or more principal metallic alloying elements – which when combined provide a high entropy of mixing [1]. Such HEAs are purported to result in single phase alloys, that may be BCC, FCC or HCP structured, depending on empirical factors outlined by Zhang [8, 11]. In practice, however, there have been demonstrations of HEAs that are multi-phase, or with a low(er) entropy of mixing, in addition to variants that either have less than 5 alloying elements or which have minor alloying additions. As a consequence, a more general description of such alloys is multi-principal element alloys (MPEAs) or compositionally complex alloys (CCAs).

Whilst a large number of CCA compositions have been studied (empirically) in the past decade, the number explored to date is on the order of ~$10^3$ alloys. Given that CCAs can have anywhere from 3 to 8+ alloying elements –inclusive of the majority of the periodic table – the number of possible CCAs is in well in excess of >>$10^{20}$ alloys – and in fact, a recent review by Cantor has elegantly indicated the number of possibilities is essentially without bounds, in a universal sense [13]. As a consequence, exploration of new alloys using methods such as generative models, is worthy of exploration. There will simply not be enough time, or effort available, for conventional (empirical) alloy design methods to be employed. Herein, a model architecture for alloy design and prediction was curated on the basis that there is (in a relative sense) only a limited empirical dataset of CCAs reported to date that can be exploited for model 'training', precluding a conventional machine learning approach. It is noted that this is a dynamic area of research, with recent studies also employing machine learning for the exploration of HEAs more generally [14], refractory HEAs [15] or specifically, HEAs with a desired HCP structure [16]. In the current work, we provide a model that is a 'whole system' from concept, to validation.

*1.1 Generative models*

Generative models are models that perform unsupervised learning and capable of generating new samples from the same distribution of training data. The present state-of-the-art, generative adversarial networks (GANs), which were proposed by Goodfellow in 2014 [17], have been widely applied in areas such as image generation [18-20], image style transfer [21], and image inpainting [22]. GANs have been demonstrated to outperform other generative models in the field of image analysis [23]. Compared to other generative models such as Boltzmann machines and general stochastic networks (GSNs), GANs do not require the implementation of Markov chain approximation that are difficult to optimise and exhibit low efficiency when dealing with data in high-dimensional spaces [23]. Consequently, GANs have also been recently applied in the fields of biology, pharmacology and materials science. In 2019, Nouira et al. developed CrystalGAN which generated novel chemically stable multi-element crystallographic structures applied in the development of new hydrogen storage materials [24]. More recently, Méndez-Lucio et al. applied Wasserstein GANs with gradient penalty (WGANs-GP) [25, 26] in drug discovery to generate candidate molecules that have a high probability of inducing a desired transcriptomic profile [27].

In this work, a generative adversarial model is proposed, called compositionally complex alloy research directive inference GAN (cardiGAN). The cardiGAN may learn from features of existing literature reported CCAs and generate formulas of novel CCAs that can potentially target desirable properties and phases. The cardiGAN model is constructed based on Wasserstein GANs (WGANs) which are a kind of GAN variant that uses the Wasserstein distance as the loss function [25].

The approach employed herein involves initially training a classification neural network, a phase classifier, to estimate the mapping between the features of CCA alloy and their respective phases. The rationale for this is that the major phase of a CCA is a key design parameter; and there is desirability for single-phase CCAs, and there may be preference for a specific phase, for example face centred cubic (FCC) CCAs are known to be less brittle [28-30]. This phase "classifier", acts as a guide for the "generator" during the training process of the cardiGAN. The cardiGAN model is then trained, and the trained model is used to generate new candidate CCA compositions as the search space for novel CCAs. Based on the prescribed ranges for empirical parameters associated with HEAs (covered by Zhang [4, 8, 11] and Guo [31]) and described herein below, the cardiGAN prediction results are presented.

## 2. Method and Approach

*2.1 Training data and data representation*

The dataset used to train the model is composed of the chemical and thermodynamic information of 278 existing literature reported CCAs, the dataset comprises 3 ternary alloys, 23 quaternary alloys, 100 quinary alloys, and 152 higher order alloys with at least 6 elements. The dataset includes alloys element compositions, empirical parameters and their reported phases. Based on the reported phases, the 278 alloys are divided into three classes: 101 single solid-solution alloys, 79 alloys with mixed solid-solution phases, and 98 alloys with secondary phases (the models and dataset can be obtained at https://github.com/anucecszl/cardiGAN).

The elements explored to predict novel CCAs are the 56 elements shown in Table 1. It is noted that numerous metals were excluded on the basis of impracticality, such as those that are radioactive, toxic, or highly reactive, and several non-metal elements are included because of their existence in the training set. The elemental composition of the alloy formulas in the training dataset are represented as 56-dimensional vectors, $\mathbf{c} = [m_1, m_2, \ldots, m_{55}, m_{56}]^T$ where $\Sigma_{i=1}^{56} m_i = 1$, with each dimension corresponding to the molar ratio of a specific element.

There are several 'empirical parameters' in the literature that have been used for HEA phase formation prediction, such as the enthalpy of mixing, entropy of mixing, atomic size difference, and omega (Ω) calculation [4, 8, 9, 32]. The empirical parameters presented in Table 2 are calculated and combined with the composition vectors as inputs for training.

**Table 1.** Elements explored to predict novel CCAs

| Ag | Al | Au | B  | Be | Bi | C  | Ca |
|----|----|----|----|----|----|----|----|
| Cd | Ce | Co | Cr | Cu | Dy | Er | Fe |
| Gd | Ge | Hf | Ho | In | Ir | La | Li |
| Lu | Mg | Mn | Mo | Nb | Nd | Ni | Os |
| P  | Pb | Pd | Pr | Pt | Re | Rh | Ru |
| Sb | Sc | Si | Sm | Sn | Sr | Ta | Tb |
| Ti | Tm | V  | W  | Y  | Yb | Zn | Zr |

**Table 2.** The calculation formulas of 12 empirical parameters utilized as training inputs herein.

| Empirical Parameter | Formula |
| --- | --- |
| Enthalpy of mixing | $\Delta H_{mix} = \Sigma_{i=1, i \neq j}^{n} 4H_{i,j} c_i c_j$ |
| Atomic size difference | $\delta = 100 \times \sqrt{\Sigma_{i=1}^{n} c_i (1 - r_i/a)^2}$ |
| $\Omega$ | $\Omega = T_m \Delta S_{mix} / |\Delta H_{mix}|$ |
| Entropy of mixing | $\Delta S_{mix} = -R \Sigma_{i=1}^{n} c_i \ln(c_i)$ |
| Average melting temperature | $T_m = \Sigma_{i=1}^{n} c_i T_i$ |
| Valence electron concentration | $VEC = \Sigma_{i=1}^{n} c_i VEC_i$ |
| Standard deviation of electronegativity | $\Delta X = 100 \times \sqrt{\Sigma_{i=1}^{n} c_i (1 - X_i/X)^2}$ |
| Standard deviation of enthalpy | $\sigma_{\Delta H} = \sqrt{\Sigma_{i=1, i \neq j}^{n} c_i c_j (H_{i,j} - \Delta H_{mix})^2}$ |
| Average atomic radii | $a = \Sigma_{i=1}^{n} c_i r_i$ |
| Standard deviation of melting temperature | $\sigma_{T_m} = 100 \times \sqrt{\Sigma_{i=1}^{n} c_i (1 - T_i/T_m)^2}$ |
| Standard deviation of valence electron concentration | $\sigma_{VEC} = \sqrt{\Sigma_{i=1}^{n} c_i (VEC_i - VEC)^2}$ |
| Electronegativity | $X = \Sigma_{i=1}^{n} c_i X_i$ |

$c_i$: The molar ratio of the $i_{th}$ element in the alloy composition.

$r_i$: The atomic radius of the $i_{th}$ element in the alloy composition.

$R$: The ideal gas constant.

*2.2 Model configuration*

As illustrated in **Fig. 1**, the cardiGAN model has four principal components: a **generator** to synthetically generate CCA formulas, a **discriminator** (serving the function of "critic") to calculate the distance between the distributions of the synthetically generated and training data, a **phase classifier** (serving as the "guide" or "regularizer" of the GAN model) to predict the phases of the generated candidate alloys, and a **parameter calculator** to calculate the 12 empirical parameters for the CCA formulas.

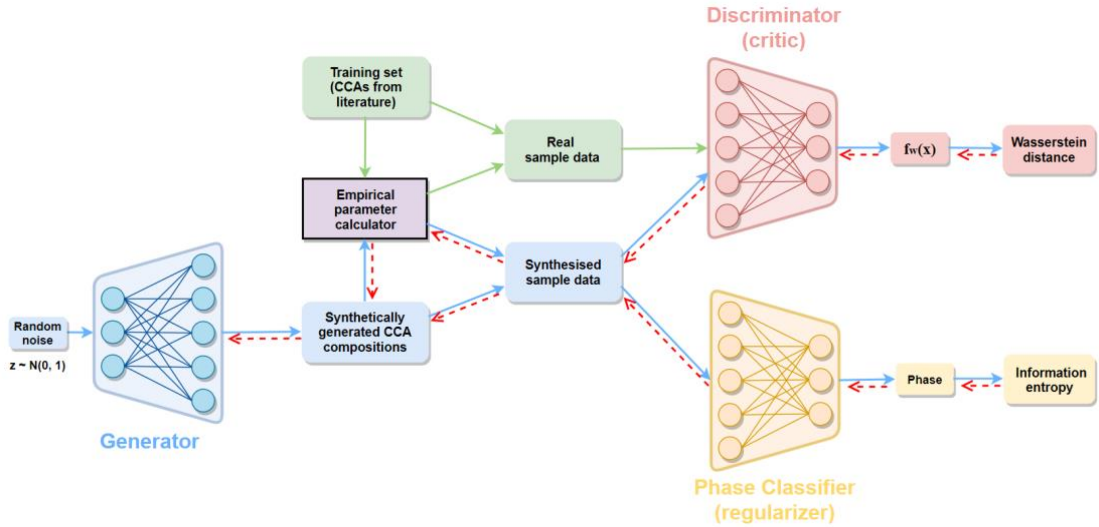

**Figure 1.** Configuration of the cardiGAN model employed herein. The red dashed arrows indicate the direction of error backpropagation.

The **generator** and the **discriminator** compose a WGAN, which is a GAN variant proposed by Arjovsky in 2017 [25]. Unlike vanilla (i.e. typical) GANs, the discriminator in the WGAN does not output a probability of the input data being real or fake; instead the output of the discriminator $f_w(x)$ is a scalar value used to calculate the Wasserstein distance between the generated and training set distributions which serves as the loss function. If one considers two distributions as different ways of placing a group of boxes, then the Wasserstein distance is the minimum cost of transforming one distribution into the other [33]. The Wasserstein distance loss metric provides two advantages:

(1) It provides a smooth gradient during training, which aids in stabilising the training process and overcomes several common GAN issues such as non-convergence, diminished gradient, mode collapse, etc. [25, 26]

(2) The Wasserstein distance correlates with the performance of the generator and the quality of the synthetically generated compositions, so one may directly monitor the model performance via interpretation of the loss [25].

*2.3 Formula used to calculate Wasserstein distance*

According to the Kantorovich-Rubinstein duality [33, 34] the Wasserstein distance between two distributions can be calculated according to the following function below [25]:

$$W(P_r, P_g) = \frac{1}{K} \sup_{\|f\|_L \leq K} E_{x \sim P_r}[f_w(x)] - E_{x \sim P_g}[f_w(x)] \qquad (1)$$

where $x$ is the input of the discriminator, $P_r$ and $P_g$ represent the distributions of the training and generated dataset, the supremum is over all the K-Lipschitz functions [35] $f_w: x \to R$. Because the Wasserstein distance is intractable, instead of directly calculating it, we train the discriminator (critic) to estimate a K-Lipschitz function $f_w$ that could maximize the value of $E_{x \sim P_r}[f_w(x)] - E_{x \sim P_g}[f_w(x)]$ whose maximum approximates the Wasserstein distance between the generated and training distributions [25]. At the same time, the generator is trained to estimate a continuous density function of $P_g$ that could minimize $E_{x \sim P_r}[f_w(x)] - E_{x \sim P_g}[f_w(x)]$. Given the discriminator and the generator are trained in opposite directions, the training process of the WGAN becomes adversarial. As the generator is trained to minimize the Wasserstein distance, the generator gradually fits the distribution of the training data until reaching equilibrium. Since there are only 278 discrete data in the training set and the fitted generator distribution is continuous, the fitted distribution includes numerous novel data which comprise our search space for novel CCAs.

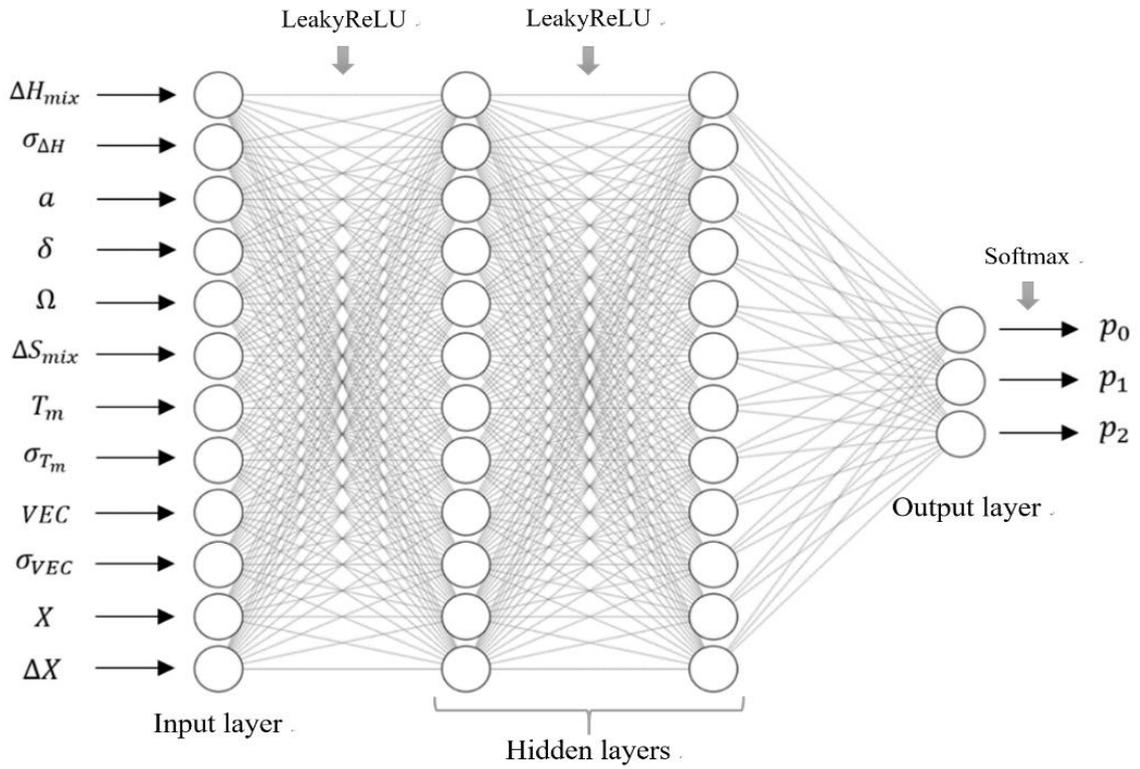

**Figure 2.** The architecture of the phase classifier in the present study

*2.4 Phase classifier and its functionality*

The **phase classifier** is a classification neural network (depicted in Figure 2) to predict the probabilities of a given alloy's 12 empirical parameters forming each of three phase states: single solid-solution phase, mixed solid-solution phase, and solid-solution phase with secondary phases. The phase classifier is pre-trained on the 12 empirical parameters and reported phases of the 278 CCAs in the training set. Herein, the trained phase classifier achieved a prediction accuracy of 80%, which is acceptable given the fact that, besides the empirical parameters, the formation of phases also relies on other factors such as alloy production method, cooling rate during production and thermomechanical history of alloys, which introduces noise into the training set and has an impact upon prediction accuracy.

The phase classifier could capture the common characteristics (based upon the empirical formulas) of the CCAs in each class. Due to the intractability of neural networks, and that the empirical formulas themselves are 'user generated' as opposed to theoretical, it is not possible to mechanistically determine what precisely these common characteristics are – however from a correlation analysis shown in Table 3, one may observe that some empirical parameters have exceptional correlation with the CCA phases.

**Table 3.** The Pearson correlation coefficients between empirical parameters and phase formation from 278 input alloys

| Empirical Parameter | Symbol | Phases in compositionally complex alloys | | |
| --- | --- | --- | --- | --- |
| | | Single solid-solution | Solid-solution with secondary phases | Mixed-solid solution |
| Absolute value of mixing enthalpy | $\|\Delta H_{mix}\|$ | -0.2892 | 0.3614 | -0.0743 |
| Standard deviation of melting temperature | $\sigma_{T_m}$ | -0.2894 | 0.261 | 0.0322 |
| Atomic size difference | $\delta$ | -0.1392 | 0.2233 | -0.088 |
| Standard deviation of enthalpy | $\sigma_{\Delta H}$ | -0.3182 | 0.1706 | 0.1586 |
| Standard deviation of VEC | $\sigma_{VEC}$ | -0.4065 | 0.1572 | 0.2669 |
| Entropy of mixing | $\Delta S_{mix}$ | -0.2978 | 0.1174 | 0.1932 |
| Standard deviation of electronegativity | $\Delta X$ | 0.0205 | 0.0857 | -0.1127 |
| Electronegativity | $X$ | -0.1891 | 0.0309 | 0.1689 |
| Valence electron concentration | $VEC$ | -0.1832 | -0.0689 | 0.2685 |
| Unitless parameter $\Omega$ | $\Omega$ | 0.1382 | -0.077 | -0.0658 |
| Average atomic radius | $a$ | 0.3429 | -0.1611 | -0.1951 |
| Average melting temperature | $T_m$ | 0.3957 | -0.1746 | -0.237 |

*To discover the correlations between the empirical parameters and formation of the three phases, the Pearson correlation coefficients between empirical parameters and reported phases are calculated, which is a coefficient developed by Pearson that measures the covariance of the two variables [36]. The Pearson correlation coefficient varies between -1 and 1, where a '+' sign indicates a positive relationship, and a '-' sign indicates a negative relationship.*

For example, the alloys possessing single solid-solution phase have positive correlations with, melting temperature ($T_m$) and $\Omega$, negative correlations with absolute value of mixing enthalpy ($|\Delta H_{mix}|$), atomic size difference ($\delta$), standard deviation of enthalpy ($\sigma_{\Delta H}$) and standard deviation of VEC ($\sigma_{VEC}$). A small(er) $|\Delta H_{mix}|$ and $\delta$ was determined to contribute to single solid-solution phase formation - which may be a selection criteria for single-phase formation. It is noteworthy that $\sigma_{\Delta H}$ and $\sigma_{VEC}$ that are generally not used as prediction paramters also show a negative correlation with single-phasee solid solutions. It is posited that those two paramters are useful for phase formation prediction and could be more generally used in future related research. Taking $\sigma_{\Delta H}$ as an example, $|\Delta H_{mix}|$ itself may be of limited utility, as in some cases the average of a very negative binary enthalpy of mixing and a very positive binary enthalpy of mixing could be near-zero. However, in this case, the $\sigma_{\Delta H}$ is, absolutely large, and indicative of elemental segregation and/or intermatallics formation.

The phase classifier plays an important role in cardiGAN's training; while the discriminator teaches the generator how to fit a proper density function for the training data, the phase classifier regularizes the density function by enforcing candidates have distinct phase structures. During cardiGAN's training, for each batch of generated CCA formulas, the parameter calculator takes their element compositions as inputs and calculate their 12 empirical parameters. Then, based on the calculation results, the pre-trained phase classifier predicts the phases of the generated alloys. For each alloy, the phase classifier outputs three scalar values, $p_0$, $p_1$, $p_2$ with restrictions of $p_0 + p_1 + p_2 = 1$, which indicate the probabilities the given alloy belonging to one of the three classes. These outputs are used to calculate the average conditional information entropy $S$ as the equation down below

$$S = H(phase|alloy) = -\frac{1}{n}\Sigma_{i=1}^{n}\Sigma_{k=1}^{3} p_{ik} \ln(p_{ik}) \qquad (2)$$

where $p_{ik}$ is the probability of the $i_{th}$ generated alloy belonging to the $k_{th}$ class, and $n$ is the number of generated alloys.

The average conditional information entropy $S$ measures the randomness of the distribution of *phase* given the formulas of generated alloys. The less random the phase variables are, the more predictable the generated alloys. Ideally, it reaches its lower bound, 0, if and only if all the generated data can be classified into one of the three classes with 100% confidence, which means for the $i_{th}$ alloy, $p_{i0}$ or $p_{i1}$ or $p_{i2} = 1$.

*2.5 Structure of the phase classifier*

The phase classifier model is based on a classification artificial neural network (ANN) which contains one input layer for the 12 empirical parameters, two hidden layers and one output layer for the label. The hidden layers are activated by LeakyReLU activation function with predetermined slope a = 0.01. As illustrated in Figure 3, unlike Rectified Linear Units (ReLU) activation function, LeakyReLU doesn't have zero-slope parts so fixes the "dying ReLU" problem where neurons become inactive and output 0 for any input [37]. It is also more balanced than ReLU therefore helps the model learn faster [38]. The output layer of the phase classifier is activated by softmax function [39] which is an activation function commonly used in multi-class classification problem.

In this setting, the outputs of the phase classifier are calculated as:

$$p_i = \frac{e^{y_i}}{\Sigma_{j=0}^{2} e^{y_j}} \qquad (3)$$

for i = 0, 1, 2 where $y_i$ is the value in the $i_{th}$ output neuron before activation.

The phase classifier is trained to minimize the cross-entropy loss, which measures the difference between the predicted phase and real phase. The cross-entropy loss function is defined as $-\Sigma_{i=1}^{K} q_i \log(p_i)$, where $q_i$ is the real phase label of the training data, and $p_i$ is the output of the phase classifier [40].

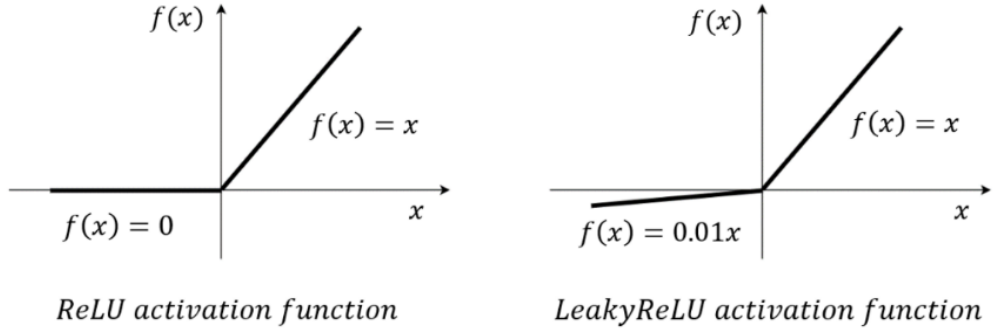

**Figure 3.** ReLU activation function vs. LeakyReLU activation function.

*2.6 Structure of cardiGAN and data flow*

*2.6.1 Generator*

The generator is a generative artificial neural network with one hidden layer. The goal of the generator is to estimate a mapping from a known distribution, such as Gaussian, to the target distribution (the distribution of training data). In our experiments, the input of the generator is a 12-dimensional Gaussian distributed noise $z \sim N(0,1)$, and the output of the generator is a 56-dimensional vector $m = [m_1, m_2, \ldots, m_{56}]^T$ with each dimension representing the molar ratio of a specific element in the generated CCA formula. Since the molar ratios of elements cannot be negative, and most existing high-entropy alloys are composed of 3 to 7 elements, the output of the generator should be a non-negative, sparse vector. To produce such output, a ReLU activation function is applied as the activation function of the output layer, which will enhance sparsity and set all the negative values in the output vector to zero.

*2.6.2 Discriminator*

The input of the discriminator includes the molar ratios of the generated and training CCA formulas along with their empirical parameters. The real CCA parameters are precalculated and can be retrieved from the training set, while the generated alloys are fed to the empirical parameter calculator to produce the 12 empirical parameters for the generated CCA formulas. The discriminator neural network is composed of two fully connected layers activated by LeakyReLU activation function and one single-neuron output layer. The input of the discriminator is a 68-dimensional vector containing the normalized molar ratios and empirical parameters of the generated and training CCA formulas, $x = [m^T, \mu^T]^T$ where $\mu = [\Delta H_{mix}, \sigma_{\Delta H}, \ldots, X, \Delta X]^T$. The output of the discriminator is a scalar value $f_w(x)$ which is a K-Lipschitz function for some constant K.

*2.6.3 Loss functions of the generator and the discriminator*

As we have mentioned before, during training, the discriminator is trained to estimate a $f_w$ that could maximize the value of the Wasserstein distance $E_{x \sim P_r}[f_w(x)] - E_{x \sim P_g}[f_w(x)]$. As described in [25], training the discriminator (critic) improves the reliability of the gradient of the Wasserstein distance. In our experiment, the discriminator is trained five times for each epoch of generator's training. On the other hand, the generator neural network is trained to fit a distribution $P_r$ that could minimize $E_{x \sim P_r}[f_w(x)] - E_{x \sim P_g}[f_w(x)]$. Because the first term is independent of the generator, the loss for the generator can be eliminated to $-E_{x \sim P_g}[f_w(x)]$. The loss functions for the generator and the discriminator, $L_g$ and $L_d$, are shown as follows:

$$L_g = - E_{x \sim P_g}[f_w(\boldsymbol{x})] \qquad (4)$$

$$L_d = E_{x \sim P_g}[f_w(\boldsymbol{x})] - E_{x \sim P_r}[f_w(\boldsymbol{x})] \qquad (5)$$

Here, it is worth mentioning that, since the generator is also trained to minimize the average information entropy $S$, the complete loss function of the generator has $S$ as the regularization term. This way, besides fitting a density function for the training data, the generator is also regularized to generate alloys whose phases are more predictable (to the phase classifier). The backpropagated gradients coming from S update the generator's weights toward a direction of generating alloys that have distinguishable empirical properties.

*2.6.4 Backpropagation of cardiGAN*

Error backpropagation is a widely applied algorithm to train NNs. Because the inputs of the discriminator are composed of the generated formulas and the calculated empirical parameters, the generator's loss $L_g$ can be divided into two parts: the error coming from the alloy formulas $L_f$ and the error coming from the empirical parameters $L_p$. As illustrated in Figure 4, when doing error backpropagation, $L_g$ is decomposed and passed to the generator through two routes. The formula loss $L_f$ is backpropagated directly to the generator through generated formulas, whereas the parameter loss $L_p$, along with the average conditional information entropy $S$, will first pass through the empirical parameter calculator, then to the generator. The two branches merge into the complete generator loss $L_g + S$ at the output layer of the generator. As suggested in [25], the weights of the generator are updated using RMSprop which is an optimization algorithm proposed by Hinton in 2012 [41].

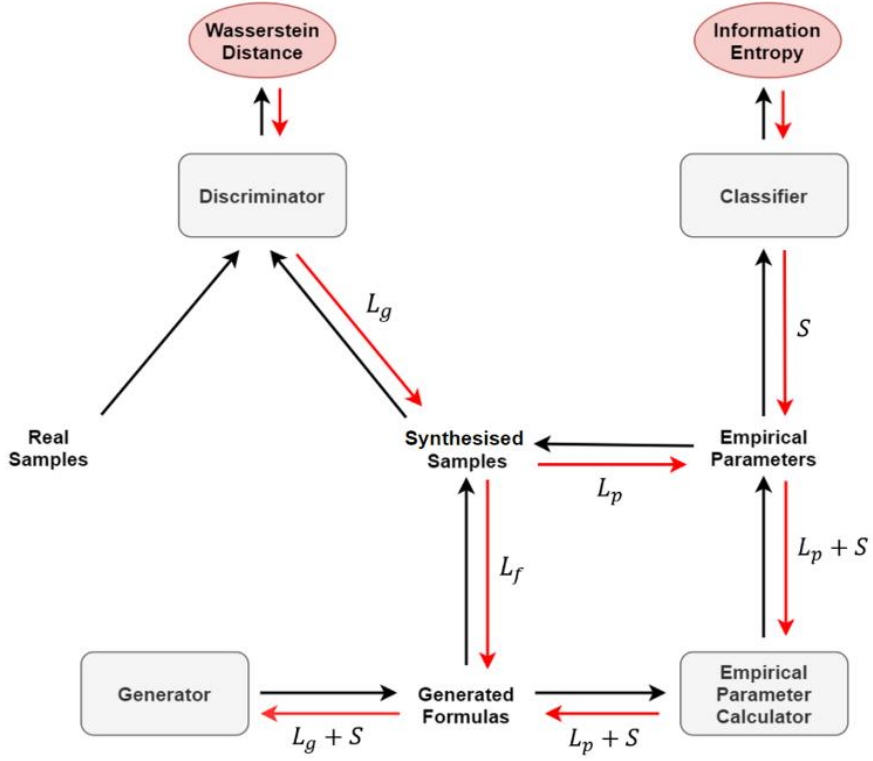

**Figure 4.** The error backpropagation of generator

*2.7 Model evaluation*

One of the most popular evaluation metrics for GANs is inception score, proposed by Salimans et al. in 2016 [42]. To evaluate a generative model, we usually assess its performance in two aspects: (1). The quality of generated samples. (2). The diversity of generated samples. In order to understand how the inception score evaluates these two aspects, we first take a look at the formula for calculation.

The formula for calculating the inception score of the generator $G$ is as follows:

$$IS(G) = exp\,[\Sigma_i\, p(x_i) D_{KL}(p(y|x_i)|p(y))] \qquad (6)$$

$x_i$: The $i_{th}$ generated sample.

$p(x_i)$: The probability distribution of $x_i$. As all the generated samples are treated even, $p(x_i) = \frac{1}{N}$.

$p(y|x_i)$: The probability (predicted by the phase classifier) of $x_i$ belongs to class $y$.

$p(y)$: The marginal probability which indicates the overall probability of the generated samples belong to class $y$, which is calculated as $p(y) = \Sigma_i\, p(x_i) p(y|x_i)$

$D_{KL}(p(y|x_i)|p(y))$: The Kullback-Leibler divergence (KL divergence) between $p(y|x_i)$ and $p(y)$, which measures the distance between the two probability distributions [43].

$$D_{KL}(p(y|\pmb{x}_i)|p(y)) = \Sigma_k\, p(y_k|\pmb{x}_i)\log\frac{p(y_k|\pmb{x}_i)}{p(y_k)} \qquad (7)$$

where $p(y_k|\pmb{x}_i)$ indicates the probability of the generated sample $\pmb{x}_i$ belongs to class $y_k$ and the marginal probability $p(y_k) = \Sigma_k\, p(y_k|\pmb{x}_i)$.

By taking the logarithm on both sides of equation (3), we could decompose the inception score into two parts with the following derivation:

$$\begin{aligned}
\log(IS(G)) &= \Sigma_{i,k}\, p(\pmb{x}_i)p(y_k|\pmb{x}_i)\log\frac{p(y_k|\pmb{x}_i)}{p(y_k)} \qquad (8)\\
&= \Sigma_{i,k}\, p(\pmb{x}_i, y_k)\log\frac{p(y_k|\pmb{x}_i)}{p(y_k)}\\
&= \Sigma_{i,k}\, p(\pmb{x}_i, y_k)\log\frac{p(y_k|\pmb{x}_i)p(\pmb{x}_i)}{p(y_k)p(\pmb{x}_i)}\\
&= \Sigma_{i,k}\, p(\pmb{x}_i, y_k)\log\frac{p(\pmb{x}_i, y_k)}{p(y_k)p(\pmb{x}_i)}\\
&= I(y;\pmb{x}) = H(y) - H(y|x)
\end{aligned}$$

where the mutual information $I(y;\pmb{x})$ measures the mutual dependence between the two variables and is equivalent to the difference between $H(y)$ and $H(y|x)$.

The information entropy $H(y)$ measures the average level of uncertainty in the distribution of $y$. It provides information about how evenly the generated samples are distributed in each class. $H(y)$ takes its maximum value $\log(K)$ when all samples could be evenly classified to each class, where K is the number of classes. And it takes its minimum value 0, when all samples are classified to the same class. So, the value of $H(y)$ is positively correlated with the diversity of generated samples. The $H(y|x)$, on the other hand, measures the average level of uncertainty of individual samples belonging to each class. Like $H(y)$, the range of $H(y|x)$ is also $[0, \log(K)]$. But, unlike our goal of increasing $H(y)$, which means increasing the diversity of generated samples, we aim to train a model that has a near-zero $H(y|x)$.

In the context of thee present study, smaller values of $H(y|x)$ indicate that the phases of the generated CCA formulas are more predictable (by the phase classifier). As samples with unpredictable or random phases are less useful, here we consider predictability as a measure of "quality" of the alloys. Given $\log(IS(G)) = H(y) - H(y|x)$, by calculating the inception score of the model, we can simultaneously assess the quality and diversity of the generated samples. And as we are more concerned about the predictability of the generated samples, we also take $H(y|x)$ as a separate indicator when evaluating the performance of the cardiGAN model.

# 3. Results and Discussion

*3.1 Evaluation Results*

To evaluate the performance of the cardiGAN model and the effectiveness of the phase classifier's regularization, 20 cardiGAN models were trained, with half of the generators trained under regularization. The inception score and conditional information entropy $H(y|x)$ were calculated from the prediction results of the generated samples, with each trained generator generating 10,000 CCA compositions.

Table 4. The evaluation results of cardiGAN

| Group | $H(y|x)$ | $H(y)$ | IS |
| --- | --- | --- | --- |
| without regularization | 0.396 | 1.08 | 1.98 |
| with regularization | 0.207 | 1.09 | 2.42 |

As shown in Table 4, with the regularization effect of phase classifier, the conditional information entropy $H(y|x)$ of the generated samples is significantly reduced from 0.396 to 0.207, which indicates that the generators trained under regularization could generate more predictable samples. However, there is no obvious improvement in the value of $H(y)$, since the generator without regularization could already evenly generate data in the three classes and achieve an $H(y) = 1.08$ which is close to the optimal value $\log 3 \approx 1.0986$. Benefit from the substantial reduction in $H(y|x)$, the inception score of the generator $IS(G) = e^{H(y)-H(y|x)}$ increased from 1.98 to 2.42. It is noted that the optimal value of $IS(G)$ is 3.

*3.2 Visualisation*

To further understand the distribution of generated CCA samples, dimension reduction algorithms such as t-distributed stochastic neighbor embedding (t-SNE) [44] and uniform manifold approximation and projection (UMAP) [45] were applied to convert the generated 56-dimensional data distribution, which contains the molar ratios of the generated CCAs, into 2-dimensional planar data distribution. As shown in Figures 5 and 6, after dimensionality reduction, data with similar features are clustered together. It's noted that, each figure is composed of the element molar ratios of 2,000 generated formulas and 278 training formulas, in which the training compositions are marked in blue, and the generated compositions are marked in red.

As can be seen from Figures 5 and 6, the distribution of synthetically generated compositions has roughly the same shape as the distribution of training (empirical) compositions. There is no obvious mode collapse in the generated distribution, and most training samples are covered (overlapped) by the generated distribution. In addition, we observed large amount of novel data in the generated distribution. Most of these novel data appear around the training data or between training data clusters, which indicates that these novel generated data may have similar element compositions to the training samples, or have hybrid element compositions from multiple samples.

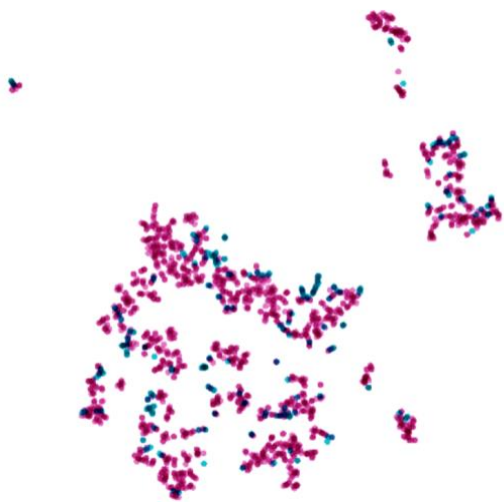

**Figure 5.** T-SNE dimensionally reduced data distribution of training (i.e. real empirical) in blue, and synthetically generated data in red, for the molar ratios of formulas.

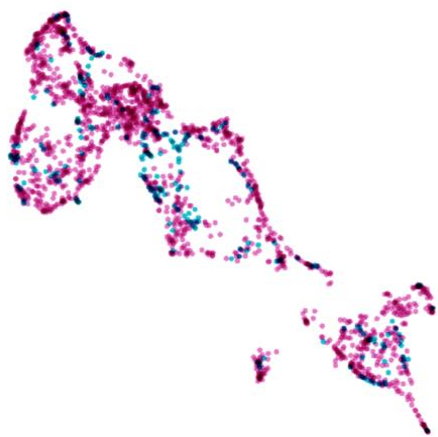

**Figure 6.** UMAP dimensionally reduced data distribution of training (i.e. real empirical) in blue, and synthetically generated data in red, for the molar ratios of formulas .

*3.3 Model utilisation*

The search space for novel single phase CCA was composed of the 200,000 candidates generated by the 20 trained generators. To narrow the search space, all the candidates with similar compositions to training samples were removed, and only those samples predicted by the phase classifier as having a high probability, ≥ 80%, of possessing single were are retained. To further increase the possibility that the selected candidate alloys are single phase, only generated samples with desired empirical parameters were kept (noting the limitation that such parameters are indeed empirically determined) . It was noted that the selection criterion for the entropy of mixing ($\Delta S_{mix}$) is the minimum value found in the training set since alloys with moderately smaller entropy values were also considered, instead of using the prescribed ranges from the literature.

From the 200,000 candidate alloys, 3,809 generated samples were selected as candidates as novel single phase CCA. From these, 188 candidates with density less than 8 g/cm³ and approximate cost less of than $USD 12/g were evaluated; thus the selection criteria are summarised in Table 5. The element cost were calculated was based on the data provided in [46].

**Table 5.** The selection criteria for single phase CCA with desired properties.

| Selection criterion | Reference |
| --- | --- |
| $-22 < \Delta H_{mix} < 7$ kJ/mol | [11, 31] |
| $\delta < 8.5\%$ | [11, 31] |
| $\Omega > 1.1$ | [4, 8] |
| $\Delta S_{mix} \geq 9.09$ J/(K.mol) | - |
| Density $< 8$ g/cm³ | - |
| Price $< 12$ USD/kg | - |

The 188 CCA candidates according to the critera in Table 5 are plotted in Figure 7, which allows us to visualise the optimized subsection, from which a candidates was chosen for production. In order to provide the reader a broader insight into the type of outputs from the cardiGAN model, Table A1 (see Appendix) reveals the nature of predicted outputs typical of cardiGAN.

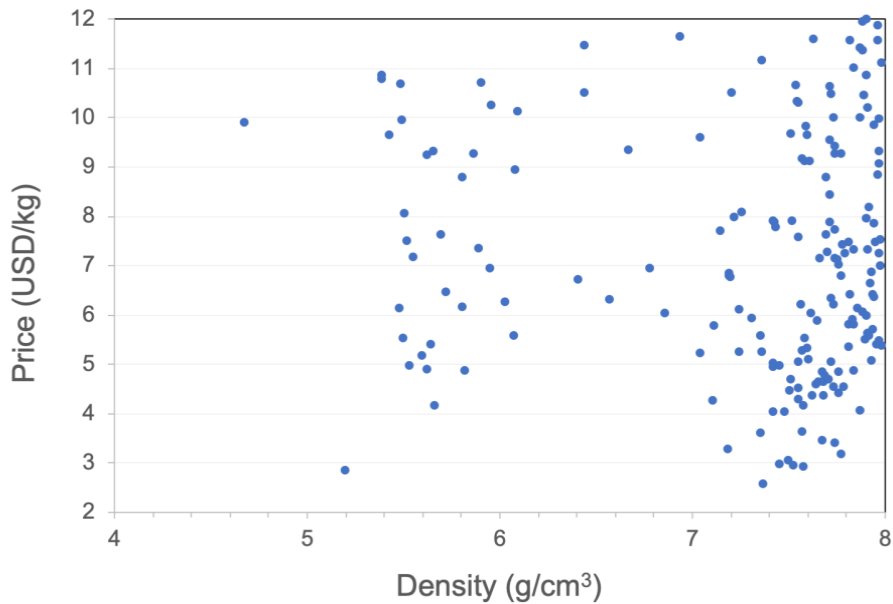

**Figure 7**. Predicted CCA candidates predicted utilising the cardiGAN model; a total of 188 candidate alloys shown according to their calculated desnisty and cost (whilst adhering to the critera in Table 5).

*3.4 Empirical validation and general discussion*

In ordet to experimentally validate the cardiGAN model, an alloy was selected for empirical production, the composition and details of which are provided in Table 6.

**Table 6.** The nominated CCA candidate prepared in this study based on cardigan prediction, and the assocaited empirical properties.

| Composition | $\Delta H_{mix}$ | $\delta$ | $\Delta S_{mix}$ | $\Omega$ | VEC | Phase* | Density | Cost |
|---|---|---|---|---|---|---|---|---|
| $Al_5Co_8Cu_{35}Fe_{19}Ni_{23}V_{11}$ | 0.81 kJ/mol | 2.25% | 13.41 J/(K.mol) | 26.99 | 9.01 | 94.8% | 7.96 g/cm$^3$ | 11.55 USD/kg |

*\* The values in the Phase column indicate the probability the alloys having a single solid solution phase*

Produced by arc-melting, the $Al_5Co_8Cu_{35}Fe_{19}Ni_{23}V_{11}$ CCA was predicted to be single-phase and to posess an FCC crystal structure (deemed potentially favourable properties such as ductility). Of all the alloys listed in Figure 7, this alloy was also selected because of its ease of production with readily available component elements. Following production of $Al_5Co_8Cu_{35}Fe_{19}Ni_{23}V_{11}$, x-ray diffraction (XRD) analysis was carried out in order to validate the cardiGAN predicted phase and structure present. Figure 8 reveals the XRD data, indicating that $Al_5Co_8Cu_{35}Fe_{19}Ni_{23}V_{11}$ appears to be comprised of a single FCC phase, with a lattice parameter similar to that of Cu (~0.36 nm).

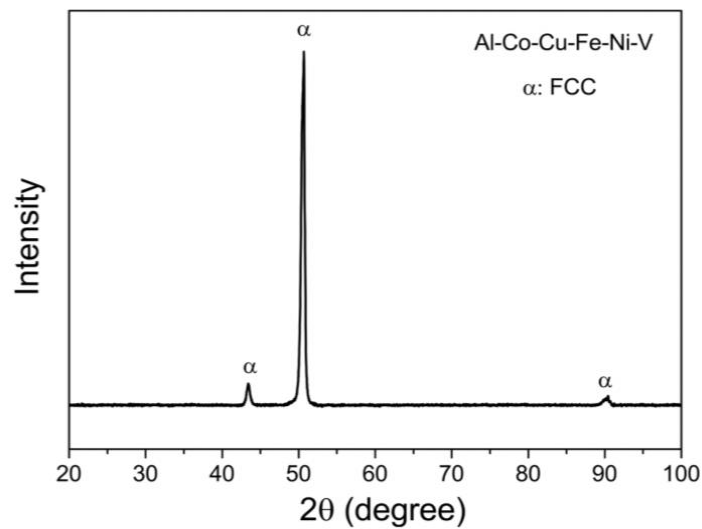

**Figure 8**. XRD pattern for $Al_5Co_8Cu_{35}Fe_{19}Ni_{23}V_{11}$.

In addition to XRD, backscattered-electron (BSE) imaging and energy-dispersive x-ray spectroscopy (EDXS) were also performed to further characterise the microstructure of $Al_5Co_8Cu_{35}Fe_{19}Ni_{23}V_{11}$, with the results are shown in Figure 9. It was readily observed from the BSE imaging that the $Al_5Co_8Cu_{35}Fe_{19}Ni_{23}V_{11}$ microstructure was not single phase. A relatively simple microstructure nonetheless, however Al$_5$Co$_8$Cu$_{35}$Fe$_{19}$Ni$_{23}$V$_{11}$ consisted of two FCC phases. For the sake of describing the alloy in more detail, the so-called 'matrix' FCC phase (deemed to be the matrix based on the phase being continuous) was rich in Co, Fe, V, Ni and Al (Figure 9b, 9c, 9e, 9f)). The non-matrix second FCC-phase was principally comprised of segregated Cu (Figure 9d). The two-phase nature of the alloy was not readily determined from XRD, due to the overlap of peaks.

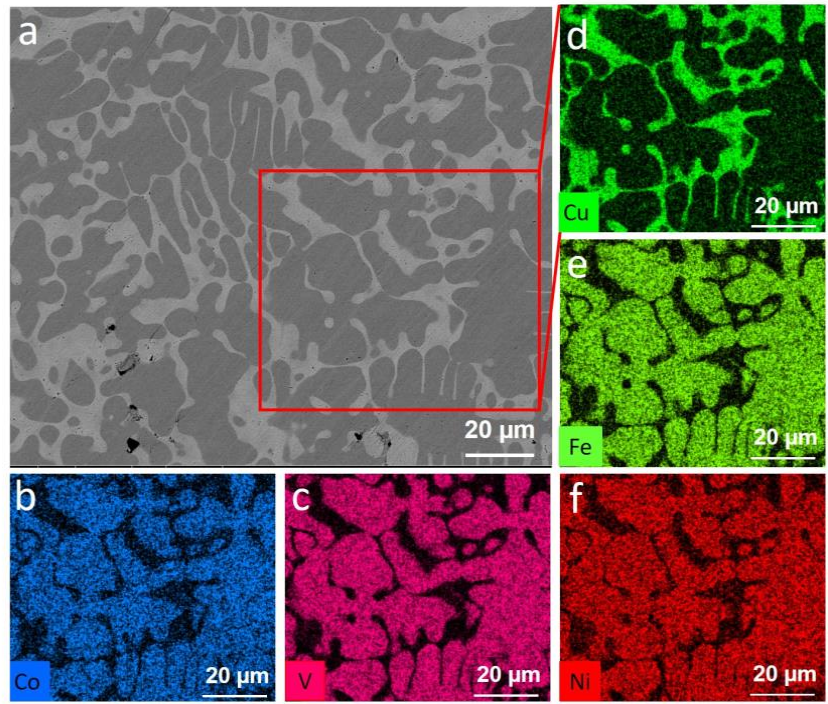

**Figure 9.** (a) BSE SEM image of Al$_5$Co$_8$Cu$_{35}$Fe$_{19}$Ni$_{23}$V$_{11}$, (b-f) the elemental distribution of Co, V, Cu, Fe and Ni. The area of EDXS mapping is indicated in (a). The map for Al was omitted, as the Al was evenly homogenously distributed throughout the alloy.

The combined development of the cardiGAN model, its predictions, and the empirical alloy (example) produced – provides some important insights that merit articulation. These facets are discussed individually.

- A generative model (GAN), specifically a WGAN, was designed, developed and deployed. This model was capable of synthetic data generation in targeted CCA prediction. The accuracy of the 'classifier' was deemed satisfactory, and the cardiGAN model was used to predict single-phase CCA compositions, one prediction of which was synthesised.

- It is apparent that the complexity of CCAs, as typified by the revelation of a two-phase microstructure for Al$_5$Co$_8$Cu$_{35}$Fe$_{19}$Ni$_{23}$V$_{11}$ could not be predicted by cardiGAN (and indeed, noting such complexity cannot be predicted by the empirical rules developed for HEA design). Although the cardiGAN prediction of FCC phase was correct, a dual FCC-phase was not anticipated – and it is clear that future work will be required to predict such scenarios. The essentialy complete segregation of Cu (as a principal element in the alloy) is, even from a metallurgical point of view, not simple to predict (followed on in the next point).

- Limitations of machine learning are perhaps anticipated, in cases where the machine learning inputs are not linked to

- physical or chemical attributes, or natural laws, specifically. This does not render machine learning of limited use, but instead is a first step to pointing to requirements for training, such as harnessing physical/chemical features, or inputs from CALPHAD (calculation of phase diagrams) in future work. However, CALPHAD alone will not be the ultimate solution, as limitations from CALPHAD based approaches are already known and articulated, including limited databases containing MPEAs, the rapid cooling rates in mst MPEA production, most MPEAs having a non-equilibrium structure or phases (in fact, even for relatively slow cooling rates or long time scales) the sluggish diffusion and so called "high entropy effect" of MPEAs often makes the attainment of equilibrium structures improbable.

- The approach presented herein has been elaborated to demonstrate that it is sound, scalable, and general. Future work is underway to make more alloys to feed into the model, and to continue 'learning' from the literature. The approach at present is meaningful, but not a silver bullet; and this is because neither cardiGAN or CALPHAD predict alloy properties. However, where machine learning (and cardiGAN) has an advantage is in the ability to permit a design-directed approach inclusive of predicting properties. It is noted that – at the time of writing – MPEA properties are only reported in a very limited subset of the already small dataset used herein, with most MPEA studies not reporting properties, and when they do, usually reporting compressive strength.

- Since the development of the cardiGAN model and the preparation of this manuscript, a number of publications that have reported new MPEA compositions, and their corresponding structures have been reported. As a validation exercise, the cardiGAN model was also used on such empirical data in order to assess its performance. The results are shown in Table 7, noting that none of the data in Table 7 was used in the development (and training) of the cardiGAN model, and hence the correlations are a true validation.

- From Table 7 it is noted that the cardiGAN model performed accurately on 13 of the 18 'new' alloys from the literature, and in the 5 cases (out of 18) where the cardiGAN model did not pe rfectly correlate, the cardiGAN model predicted additional intermetallic phases that were not present in the empricial alloys. As a consequence, a preliminary finding in terms of field performance of the cardiGAN model is that it is conservative in its selection of single phase alloy systems (which is in fact, an ideal outcome at this stage). In alloy design, one needs to start somewhere, as MPEA design extends beyond human level understanding into the so-called 'abstract', where models (be it GANs, CALPHAD, first principles, DFT, etc.) will be critical. \Empirical testing, and known 'rule of thumbs' will not be adequate. Such facets were covered elegantly in the historical recounting of the MPEA design journey of Cantor [13], which ultimately points to a requirement for open-ness to innovation. The model herein - cardiGAN - provides a democratised and simple to follow process as one discrete effort for MPEA design, that may readily integrate with other models as it evolves.

**Table 7.** Recent 'new' MPEAs reported in the literature, and a comparison of the empirical structure, with the cardiGAN predicted structure.

| Alloy | Composition | Structure from experimental study | Structure prediction from cardigan model herein | Was the cardiGAN model able to predict this outcome? | Ref. |
|---|---|---|---|---|---|
| 1 | $Al_{14}Co_{24}Cr_{22}Fe_{21}Ni_{19}$ | FCC: 62 vol. % <br> FCC/BCC: 38 vol. % | Mixed solid solution | Yes | [47] |
| 2 | $Al_{16}Co_{23}Cr_{17}Fe_{19}Ni_{25}$ | FCC: 52 vol. % <br> FCC/BCC: 48 vol. % | Mixed solid solution | Yes | [47] |
| 3 | TiZrNb (500ppm Boron) | BCC (100%) | Single solid solution | Yes | [48] |
| 4 | $NiMnFe_{0.5}Cr_{0.5}Al_{0.2}Cu_x$ ($x$ = 0.8, 1.2, and 1.6) | FCC+BCC | Mixed solid solution | Yes | [49] |
| 5 | $Ti_{45}Zr_{45}Nb_5Ta_5$ | BCC (100%) | Single solid solution | Yes | [50] |
| 6 | $Ti_{45}Zr_{45}Nb_5Mo_5$ | BCC (100%) | Single solid solution | Yes | [50] |
| 7 | $Ti_{42.5}Zr_{42.5}Nb_5Mo_5Ta_5$ | BCC (100%) | Single solid solution | Yes | [50] |
| 8 | $Cr_{30}Co_{30}Ni_{15}Ni_1C_{24}$ | FCC + HCP + Carbides | Solid solution with intermetallic compound | Yes | [51] |
| 9 | $Co_{23}Cr_{22}Cu_3Ti_{28}Fe_{24}$ | HCP | Solid solution with intermetallic compound | Partly (model predicted additional intermetallics) | [52] |
| 10 | $Ni_{38}Fe_{20}Cr_{21}Mo_6W_2Ru_{13}$ | FCC | Single solid solution | Yes | [53] |
| 11 | $Mg_{12}Al_{11}Ti_{33}Mn_{11}Nb_{33}$ | BCC | Single solid solution | Yes | [54] |
| 12 | MgVAlCrNi | BCC | Single solid solution | Yes | [55] |
| 13 | AlCuMgZn | FCC | Solid solution with intermetallic compound | Partly (model predicted additional intermetallics) | [56] |
| 14 | AlNbVTiZr | BCC (Ordered B2) | Solid solution with intermetallic compound | Yes | [57] |
| 15 | $Co_{21}Cu_{16}Fe_{21}Ti_{21}V_{21}$ | BCC (main) + FCC (secondary) | Solid solution with intermetallic compound | Yes | [58] |
| 16 | CoFeNiTiV | BCC | Solid solution with intermetallic compound | Partly (model predicted additional intermetallics) | [59] |
| 17 | FeNiTiAl | BCC | Solid solution with intermetallic compound | Partly (model predicted additional intermetallics) | [59] |
| 18 | CoFeNiTiCu | FCC | Single solid solution | Yes | [59] |

**Conclusions**

1) A generative model was developed, explored, and presented - in the context of predicting and designing CCAs. A GAN, specifically a WGAN, was preferred on the basis that there is little empirical data regarding CCAs presently available, and synthetic data generation was deemed critical in targeted CCA prediction.

2) The model presented herein, termed cardiGAN, was demonstrated as being capable of functioning with suitable accuracy in the performance of the 'classifier', and provides a significant advance in the ability to general novel CCA compositions with limited training data.

3) The cardiGAN model may be utilised to generate large numbers of candidate CCAs with minimal computational burden, generating novel compositions for which all 12 empirical parameters may be calculated (and such parameters then augmented for constraints such as density or cost, and perhaps in future, corrosion rate predictions).

4) To demonstrate the utility of cardiGAN, a novel predicted alloy was selected, and empirically produced, namely $Al_5Co_8Cu_{35}Fe_{19}Ni_{23}V_{11}$. The as-cast CCA structure of the alloys predicted was FCC, albeit a two-phase FCC (where both phases were FCC), which was unable to be predicted from the procedures in the study herein. An elaboration of discussion that followed this result was articulated, highlight a number of limitations, challenges, and opportunities for machine learning in CCA design. The cardiGAN model was also applied to new alloys in the literature, and was determined as providing high degree of correlation with its prediction of alloy structures in previously unreported alloys.

5) The cardiGAN model is a freeware tool that is available to researchers and metallurgists looking to predict and design CCAs (and therein, HEAs) in a high throughput manner.

**Acknowledgments**

The authors gratefully acknowledge the use of facilities in the Monash Centre for Electron Microscopy (MCEM) and Monash X-ray Platform (MXP).

**Author contributions**

ZL performed the majority of the model coding, with guidance from WTN. SO and RKG carried out alloy preparation, and YQ performed alloy testing. NB, WTN and ZL designed the study. NB supervised the study. All authors contributed to manuscript preparation.

**Competing interests**

The authors declare no competing interests.

**Experimental methods**

Alloy production for $Al_5Co_8Cu_{35}Fe_{19}Ni_{23}V_{11}$ was carried out using the arc melting process. Metal pieces (99.7% purity V and thee remaining elements with greater than or equal to 99.9% purity) were arc melted in a copper crucible within an inert argon environment. Each specimen was re-melted multiple times in between flips to obtain a homogenous composition, with the alloy flipped four times, respectively.

X-ray diffraction (XRD) was performed to determine the phase(s) present, with XRD data collected using a Bruker D8 Advance with a Cu target operated at 40 kV and 40 mA. The scan range was from 20° to 100°, with an increment of 0.02° at the rate of 1.2s/step. Specimens for SEM analysis were polished to a 1 μm finish using diamond paste, followed by final polishing using a 0.05 μm alumina suspension. Imaging was carried out using an FEI Nova NanoSEM 450 in backscattered electron (BSE) mode, and the instrument is fitted with a large area SDD X-ray detector allowing fast X-ray mapping.

# Appendix

**Table A1.** Compositions of cardiGAN model predicted CCAs and their associated empirical parameters. Alloys shown in alphabetical order.

| Composition | $\Delta H_{mix}$(kJ/mol) | $\sigma_{\Delta H}$(kJ/mol) | a(Å) | $\delta$(%) | $\Omega$ | $\Delta S_{mix}$(J/K.mol) | $T_m$(K) | $\sigma_{T_m}$(%) | X | $\Delta X$(%) | VEC | $\sigma_{VEC}$ | Density(g/cm$^3$) | Price(USD/g) |
|---|---|---|---|---|---|---|---|---|---|---|---|---|---|---|
| $Al_{44}Cr_2Cu_{18}Ti_{10}V_{26}$ | -13.47 | 6.96 | 1.31 | 4.40 | 1.21 | 11.11 | 1465.91 | 37.02 | 1.66 | 6.91 | 5.10 | 2.87 | 4.68 | 9.91 |
| $Al_{41}Cr_9Cu_3Fe_6Mn_{22}Zn_{19}$ | -9.00 | 5.58 | 1.33 | 5.05 | 1.65 | 12.50 | 1187.61 | 38.37 | 1.63 | 4.89 | 6.42 | 3.43 | 5.20 | 2.84 |
| $Al_{44}Co_8Cr_4Cu_{14}Fe_{16}Mn_2Ni_8Sn_3V_1$ | -8.85 | 6.47 | 1.32 | 5.02 | 2.12 | 14.16 | 1326.62 | 33.63 | 1.74 | 7.85 | 6.51 | 3.45 | 5.39 | 10.78 |
| $Al_{44}Co_9Cr_4Cu_{14}Fe_1Ni_{15}Zn_{14}$ | -9.50 | 6.31 | 1.31 | 4.04 | 1.67 | 13.13 | 1206.55 | 35.16 | 1.73 | 7.82 | 7.08 | 3.80 | 5.39 | 10.87 |
| $Al_{43}Co_6Cr_6Cu_{15}Fe_8Mn_{10}Ni_8Sn_4$ | -9.54 | 6.65 | 1.32 | 5.13 | 1.97 | 14.52 | 1294.16 | 33.11 | 1.72 | 8.39 | 6.53 | 3.49 | 5.42 | 9.64 |
| $Al_{43}Co_2Cu_{13}Fe_{14}Ni_{25}Zn_2$ | -11.86 | 6.84 | 1.31 | 4.43 | 1.31 | 11.77 | 1323.51 | 29.58 | 1.76 | 7.87 | 6.83 | 3.47 | 5.48 | 6.14 |
| $Al_{36}Co_{12}Cr_2Cu_4Fe_{10}Mn_{21}Si_2Zn_{14}$ | -13.23 | 6.84 | 1.33 | 5.27 | 1.36 | 14.29 | 1257.93 | 33.62 | 1.67 | 7.29 | 6.70 | 3.25 | 5.49 | 10.68 |
| $Al_{43}Co_4Cr_3Cu_{22}Fe_8Ni_{10}Sn_5V_5$ | -6.79 | 6.39 | 1.32 | 4.80 | 2.62 | 13.80 | 1288.52 | 34.78 | 1.75 | 8.14 | 6.85 | 3.78 | 5.50 | 9.96 |
| $Al_{38}Co_1Cr_{17}Cu_{13}Fe_{15}Ni_{13}Ti_2Zn_1$ | -10.52 | 6.86 | 1.33 | 4.89 | 1.92 | 13.80 | 1460.53 | 33.27 | 1.73 | 7.50 | 6.35 | 3.12 | 5.50 | 5.51 |
| $Al_{43}Co_3Cu_{15}Fe_9Nb_2Ni_{20}Zn_8$ | -10.87 | 6.84 | 1.31 | 4.42 | 1.52 | 12.99 | 1275.19 | 34.34 | 1.74 | 7.95 | 6.98 | 3.66 | 5.50 | 8.05 |
| $Al_{43}Co_3Cu_{19}Fe_2Ni_{23}Zn_9$ | -10.24 | 6.60 | 1.31 | 3.89 | 1.42 | 11.87 | 1220.74 | 30.88 | 1.75 | 8.13 | 7.27 | 3.80 | 5.52 | 7.50 |
| $Al_{42}Cu_{14}Fe_9Ni_{26}Zn_9$ | -10.92 | 6.75 | 1.31 | 4.20 | 1.36 | 11.82 | 1258.76 | 31.63 | 1.75 | 7.97 | 7.17 | 3.68 | 5.53 | 4.97 |
| $Al_{43}Cr_5Cu_{14}Fe_{13}Ni_{19}Sn_7$ | -8.28 | 6.81 | 1.32 | 5.14 | 2.00 | 12.86 | 1287.31 | 35.04 | 1.76 | 8.10 | 6.97 | 3.77 | 5.55 | 7.16 |
| $Al_{41}Cu_{11}Fe_{15}Ni_{31}$ | -13.43 | 6.86 | 1.32 | 4.41 | 1.11 | 10.94 | 1361.56 | 28.83 | 1.77 | 7.86 | 6.92 | 3.41 | 5.60 | 5.16 |
| $Al_{43}Cu_{19}Fe_8Ni_{16}Sn_{12}V_1$ | -5.67 | 6.52 | 1.32 | 5.37 | 2.57 | 12.38 | 1175.25 | 36.31 | 1.77 | 8.44 | 7.44 | 4.17 | 5.63 | 9.24 |
| $Al_{40}Cr_5Cu_{20}Fe_{12}Ni_{19}Zn_3$ | -8.22 | 6.65 | 1.32 | 4.48 | 2.05 | 12.59 | 1340.35 | 30.90 | 1.76 | 7.80 | 6.96 | 3.50 | 5.63 | 4.88 |
| $Al_{40}Cr_7Cu_{26}Fe_1Ni_{18}Zn_8$ | -6.72 | 6.09 | 1.31 | 4.10 | 2.26 | 12.08 | 1256.61 | 32.68 | 1.75 | 8.04 | 7.30 | 3.77 | 5.64 | 5.40 |
| $Al_{43}Cu_{14}Fe_8Ni_{23}Sn_{12}$ | -8.18 | 6.95 | 1.32 | 5.36 | 1.79 | 12.21 | 1198.08 | 36.74 | 1.78 | 8.43 | 7.45 | 4.12 | 5.66 | 9.30 |
| $Al_{38}Cu_{14}Fe_{11}Mn_{19}Ni_{18}$ | -13.46 | 6.98 | 1.33 | 4.78 | 1.25 | 12.52 | 1343.29 | 25.85 | 1.72 | 8.74 | 6.68 | 3.15 | 5.67 | 4.16 |
| $Al_{38}Co_4Cu_5Fe_{12}Ni_{27}Zn_{13}$ | -12.90 | 6.68 | 1.32 | 4.23 | 1.27 | 12.79 | 1280.87 | 34.05 | 1.75 | 7.74 | 7.35 | 3.58 | 5.69 | 7.61 |
| $Al_{38}Cr_{18}Cu_{16}Ni_{27}$ | -11.35 | 6.83 | 1.32 | 4.42 | 1.45 | 11.37 | 1443.32 | 32.53 | 1.75 | 8.00 | 6.80 | 3.37 | 5.72 | 6.46 |
| $Al_{32}Cr_{17}Cu_{11}Fe_{15}Mn_2Ni_{16}Ti_1V_6$ | -11.57 | 6.91 | 1.34 | 4.64 | 1.98 | 14.75 | 1550.26 | 31.21 | 1.73 | 7.51 | 6.48 | 2.95 | 5.80 | 6.15 |
| $Al_{38}Co_2Cr_6Cu_{18}Fe_{11}Ni_{18}Sn_7$ | -6.99 | 6.82 | 1.33 | 5.00 | 2.58 | 13.78 | 1308.24 | 34.70 | 1.77 | 8.02 | 7.29 | 3.76 | 5.80 | 8.78 |
| $Al_{34}Cr_{11}Cu_7Fe_{11}Mn_{12}Ni_{22}Zn_2$ | -13.53 | 6.85 | 1.33 | 4.73 | 1.53 | 14.30 | 1444.58 | 30.61 | 1.72 | 8.18 | 6.71 | 3.05 | 5.82 | 4.86 |
| $Al_{38}Co_5Cu_{26}Fe_7Ni_{21}Sn_1V_2$ | -8.58 | 6.86 | 1.32 | 4.16 | 1.92 | 12.40 | 1324.97 | 28.10 | 1.78 | 7.91 | 7.37 | 3.61 | 5.87 | 9.25 |

| Composition | | | | | | | | | | | | | |
|---|---|---|---|---|---|---|---|---|---|---|---|---|---|
| $Al_{36}Cr_6Cu_{19}Fe_{10}Ni_{22}Sn_5V_2$ | -8.06 | 6.97 | 1.33 | 4.78 | 2.30 | 13.66 | 1355.32 | 33.57 | 1.77 | 7.96 | 7.30 | 3.65 | 5.89 | 7.35 |
| $Al_{36}Co_9Cr_5Cu_{27}Fe_4Mn_4Ni_{11}Zn_4$ | -7.08 | 6.39 | 1.32 | 4.18 | 2.65 | 14.28 | 1313.95 | 29.84 | 1.75 | 8.09 | 7.34 | 3.54 | 5.90 | 10.69 |
| $Al_{34}Co_3Cu_{23}Fe_7Ni_{18}Ti_1Zn_{13}$ | -7.74 | 6.61 | 1.32 | 3.95 | 2.17 | 13.55 | 1241.30 | 31.79 | 1.76 | 7.90 | 7.84 | 3.70 | 5.95 | 6.94 |
| $Al_{39}Cu_{26}Ni_{21}Sn_{15}$ | -5.08 | 6.27 | 1.33 | 5.22 | 2.49 | 11.06 | 1143.44 | 35.02 | 1.80 | 8.41 | 8.13 | 4.26 | 5.96 | 10.26 |
| $Al_{33}Cu_{33}Fe_2Ni_{15}V_4Zn_{12}$ | -5.53 | 5.54 | 1.32 | 3.74 | 2.81 | 12.64 | 1228.66 | 31.50 | 1.76 | 8.11 | 7.99 | 3.79 | 6.03 | 6.26 |
| $Al_{33}Cr_7Cu_{26}Ni_{21}Zn_{12}$ | -6.39 | 6.15 | 1.32 | 3.89 | 2.53 | 12.68 | 1273.46 | 33.53 | 1.76 | 7.93 | 7.92 | 3.70 | 6.08 | 5.57 |
| $Al_{32}Co_5Cr_7Cu_7Fe_{24}Nb_2Ni_{24}$ | -13.21 | 6.79 | 1.34 | 4.69 | 1.56 | 13.56 | 1521.78 | 29.54 | 1.77 | 7.32 | 6.94 | 2.98 | 6.08 | 8.93 |
| $Al_{32}Co_8Cr_6Cu_{15}Fe_{11}Ni_{22}Zn_5$ | -9.57 | 6.93 | 1.33 | 4.21 | 2.14 | 14.61 | 1400.53 | 31.00 | 1.77 | 7.58 | 7.45 | 3.34 | 6.10 | 10.12 |
| $Al_{19}Fe_1Mn_{21}Ni_{27}V_{12}Zn_{20}$ | -16.40 | 6.17 | 1.34 | 3.68 | 1.13 | 13.48 | 1379.90 | 35.95 | 1.69 | 8.21 | 7.78 | 3.21 | 6.41 | 6.72 |
| $Al_{28}Co_{10}Cu_{16}Fe_8Mn_1Ni_{28}Zn_{10}$ | -9.66 | 6.91 | 1.33 | 3.75 | 2.01 | 14.33 | 1357.16 | 29.98 | 1.79 | 7.60 | 8.11 | 3.34 | 6.44 | 11.48 |
| $Al_{22}Co_6Fe_{15}Mn_{10}Ni_{32}V_{10}Zn_4$ | -17.61 | 6.83 | 1.34 | 3.93 | 1.30 | 14.79 | 1545.77 | 27.39 | 1.75 | 8.20 | 7.35 | 2.87 | 6.44 | 10.50 |
| $Al_{24}Cr_{13}Fe_{25}Mo_4Ni_{33}$ | -14.35 | 5.88 | 1.35 | 4.82 | 1.47 | 12.59 | 1672.80 | 29.04 | 1.80 | 8.10 | 7.09 | 2.68 | 6.57 | 6.31 |
| $Al_3Cr_{28}Mn_{39}Mo_1Nb_5Ti_{13}V_{10}$ | -5.12 | 3.82 | 1.39 | 2.40 | 4.72 | 12.82 | 1888.47 | 21.63 | 1.60 | 5.20 | 5.88 | 1.16 | 6.67 | 9.35 |
| $Al_{18}Cr_{40}Fe_{28}Mo_{13}V_1$ | -6.58 | 2.94 | 1.38 | 4.57 | 3.28 | 11.15 | 1940.31 | 29.52 | 1.76 | 9.81 | 6.01 | 1.66 | 6.78 | 6.95 |
| $Al_{21}Cr_{11}Cu_6Fe_{27}Mo_5Ni_{28}$ | -9.75 | 6.06 | 1.35 | 4.42 | 2.29 | 13.37 | 1673.54 | 28.90 | 1.81 | 7.98 | 7.39 | 2.69 | 6.86 | 6.02 |
| $Al_{25}Cr_{22}Fe_4Mo_{40}Ti_9$ | -8.88 | 4.58 | 1.38 | 5.80 | 2.74 | 11.50 | 2116.26 | 36.88 | 1.84 | 14.32 | 5.13 | 1.43 | 6.94 | 11.63 |
| $Cr_{28}Fe_{17}Mn_{37}Ti_7V_9$ | -2.36 | 2.59 | 1.40 | 1.04 | 9.40 | 11.93 | 1854.66 | 15.60 | 1.64 | 6.14 | 6.49 | 1.09 | 7.04 | 5.22 |
| $Cr_{41}Cu_5Mn_{28}V_{26}$ | 1.17 | 2.00 | 1.38 | 1.66 | 16.86 | 10.13 | 1951.54 | 16.65 | 1.63 | 4.60 | 6.26 | 1.28 | 7.04 | 9.59 |
| $Al_4Cr_{34}Cu_6Fe_{17}Mn_{35}Ti_3Zr_1$ | -1.56 | 4.52 | 1.39 | 2.61 | 14.06 | 12.33 | 1779.02 | 19.62 | 1.65 | 7.16 | 6.75 | 1.54 | 7.11 | 4.27 |
| $Al_{14}Cr_{28}Fe_{44}Mo_{13}$ | -5.77 | 2.72 | 1.38 | 4.17 | 3.51 | 10.50 | 1930.77 | 27.80 | 1.79 | 9.33 | 6.45 | 1.69 | 7.11 | 5.77 |
| $Al_{10}Cr_{34}Fe_{23}Mn_6Mo_{19}Ti_9$ | -7.08 | 4.38 | 1.39 | 3.65 | 3.91 | 13.49 | 2050.99 | 26.20 | 1.77 | 11.76 | 6.05 | 1.48 | 7.15 | 7.71 |
| $Cr_{28}Fe_{22}Mn_{42}Ti_7$ | -1.82 | 2.70 | 1.40 | 0.00 | 10.26 | 10.33 | 1804.96 | 15.31 | 1.64 | 6.72 | 6.72 | 1.03 | 7.19 | 3.27 |
| $Cr_{19}Fe_{24}Mn_{41}Nb_6Ti_{10}$ | -4.77 | 3.58 | 1.40 | 0.86 | 4.57 | 11.88 | 1833.35 | 18.63 | 1.64 | 7.00 | 6.62 | 1.21 | 7.19 | 6.83 |
| $Cr_{28}Fe_{21}Mn_{33}V_{19}$ | -1.28 | 1.55 | 1.39 | 1.41 | 16.72 | 11.33 | 1887.62 | 15.50 | 1.65 | 6.09 | 6.56 | 1.02 | 7.19 | 6.80 |
| $Cr_{31}Fe_{23}Mn_{31}Ti_2V_6Zr_7$ | -4.84 | 4.32 | 1.41 | 2.93 | 4.81 | 12.37 | 1882.75 | 14.97 | 1.64 | 8.20 | 6.53 | 1.17 | 7.20 | 6.75 |
| $Cr_{23}Fe_{22}Mn_{26}Nb_5V_{24}$ | -3.05 | 2.30 | 1.39 | 1.80 | 8.02 | 12.51 | 1952.76 | 16.79 | 1.66 | 5.97 | 6.41 | 1.12 | 7.21 | 10.51 |
| $Cr_{27}Fe_{23}Mn_{25}Mo_2V_{23}$ | -1.86 | 1.73 | 1.39 | 1.64 | 12.69 | 12.11 | 1944.20 | 15.98 | 1.67 | 7.29 | 6.48 | 1.08 | 7.22 | 7.98 |
| $Al_1Cr_{20}Fe_{28}Mn_{37}Nb_4Ti_5V_4$ | -3.90 | 3.44 | 1.40 | 1.65 | 5.86 | 12.55 | 1821.70 | 18.46 | 1.66 | 7.04 | 6.72 | 1.18 | 7.25 | 6.09 |

| Composition | | | | | | | | | | | | | | |
|---|---|---|---|---|---|---|---|---|---|---|---|---|---|---|
| $Al_6Cr_{22}Fe_{25}Mn_{35}Mo_6V_5$ | -3.03 | 3.56 | 1.39 | 2.89 | 7.77 | 12.89 | 1823.64 | 23.77 | 1.69 | 9.56 | 6.61 | 1.26 | 7.25 | 5.24 |
| $Cr_{23}Fe_7Mn_{38}Mo_{10}Ti_8V_{14}$ | -1.47 | 2.52 | 1.40 | 1.74 | 17.76 | 13.31 | 1955.73 | 21.72 | 1.67 | 10.85 | 6.23 | 1.03 | 7.26 | 8.07 |
| $Cr_{28}Cu_1Fe_{18}Mn_{42}V_7Zr_4$ | -1.90 | 3.45 | 1.40 | 2.37 | 11.17 | 11.64 | 1825.78 | 16.39 | 1.63 | 7.42 | 6.69 | 1.09 | 7.31 | 5.92 |
| $Co_3Cr_{26}Fe_{27}Mn_{37}Ti_4V_3$ | -1.84 | 2.51 | 1.40 | 0.85 | 11.50 | 11.68 | 1810.42 | 14.71 | 1.67 | 7.24 | 6.92 | 1.05 | 7.35 | 5.59 |
| $Cr_{37}Cu_1Fe_{25}Mn_{34}Ti_3$ | -0.08 | 2.29 | 1.40 | 0.39 | 233.73 | 10.32 | 1845.03 | 15.41 | 1.66 | 6.71 | 6.84 | 1.03 | 7.36 | 3.62 |
| $Al_{25}Cr_{17}Fe_{17}Mo_{42}$ | -6.26 | 2.56 | 1.38 | 5.75 | 3.72 | 11.04 | 2107.78 | 37.19 | 1.88 | 12.93 | 5.59 | 1.65 | 7.36 | 11.15 |
| $Cr_{27}Fe_{25}Mn_{35}Ni_2V_{11}$ | -1.31 | 1.83 | 1.39 | 1.21 | 16.31 | 11.60 | 1847.26 | 15.34 | 1.67 | 6.83 | 6.84 | 1.06 | 7.36 | 5.26 |
| $Al_3Cr_{22}Cu_2Fe_{37}Mn_{35}$ | -0.75 | 2.94 | 1.39 | 1.97 | 24.25 | 10.45 | 1751.48 | 16.73 | 1.69 | 7.38 | 7.10 | 1.19 | 7.37 | 2.56 |
| $Cr_{42}Fe_{36}Mn_9Mo_6Ti_7$ | -3.05 | 2.81 | 1.40 | 0.84 | 7.15 | 10.85 | 2010.24 | 15.42 | 1.73 | 8.58 | 6.68 | 1.16 | 7.42 | 5.02 |
| $Al_4Cr_{43}Fe_{35}Mn_{12}Mo_5$ | -1.99 | 2.31 | 1.40 | 2.28 | 10.54 | 10.74 | 1956.47 | 19.12 | 1.73 | 8.14 | 6.71 | 1.19 | 7.42 | 4.94 |
| $Cr_{30}Fe_{23}Mn_{37}Ni_5Ti_2V_2$ | -1.90 | 2.60 | 1.40 | 0.94 | 11.06 | 11.59 | 1817.86 | 15.14 | 1.67 | 7.31 | 6.99 | 1.14 | 7.42 | 4.03 |
| $Co_2Cr_{36}Cu_6Fe_{40}V_{15}$ | -0.49 | 3.45 | 1.39 | 1.53 | 42.94 | 10.67 | 1970.65 | 12.14 | 1.74 | 5.57 | 7.04 | 1.57 | 7.42 | 7.89 |
| $Cr_{30}Fe_{32}Mn_{21}Mo_4Zr_{13}$ | -7.59 | 5.58 | 1.42 | 3.55 | 3.09 | 12.04 | 1944.32 | 16.41 | 1.67 | 11.38 | 6.60 | 1.29 | 7.43 | 7.88 |
| $Cr_{31}Fe_8Mn_{43}V_{16}W_2$ | 0.09 | 1.32 | 1.39 | 1.37 | 220.42 | 10.76 | 1887.44 | 20.75 | 1.63 | 7.55 | 6.44 | 0.86 | 7.43 | 7.77 |
| $Cr_{31}Fe_{44}Mn_{16}V_8$ | -1.52 | 1.42 | 1.40 | 1.02 | 13.12 | 10.48 | 1909.59 | 12.51 | 1.72 | 6.34 | 6.99 | 1.05 | 7.45 | 4.97 |
| $Cr_3Cu_{17}Fe_{33}Mn_{28}Ni_7Ti_{11}$ | -2.08 | 5.95 | 1.39 | 1.53 | 10.38 | 12.93 | 1669.20 | 12.74 | 1.73 | 8.95 | 7.86 | 1.97 | 7.46 | 2.96 |
| $Cr_{20}Cu_3Fe_{31}Mn_{39}V_7$ | 0.67 | 2.14 | 1.39 | 1.11 | 30.00 | 11.30 | 1782.49 | 15.50 | 1.68 | 7.40 | 7.10 | 1.16 | 7.48 | 4.04 |
| $Cr_{28}Fe_{28}Mn_{42}$ | 0.53 | 0.77 | 1.40 | 0.28 | 31.09 | 9.24 | 1792.76 | 15.20 | 1.66 | 7.01 | 7.00 | 0.77 | 7.50 | 3.06 |
| $Cr_{31}Fe_{23}Mn_{44}Nb_2$ | 0.27 | 1.44 | 1.40 | 0.48 | 62.18 | 9.42 | 1812.83 | 17.11 | 1.65 | 6.61 | 6.88 | 0.77 | 7.51 | 4.47 |
| $Cu_{32}Mn_{21}Mo_{16}Ti_{17}V_{10}Zr_4$ | 0.60 | 6.10 | 1.39 | 3.43 | 42.55 | 13.68 | 1851.51 | 29.13 | 1.76 | 13.74 | 7.28 | 2.74 | 7.51 | 9.66 |
| $Cr_{42}Fe_{15}Mn_{35}Ni_8$ | -1.04 | 2.02 | 1.40 | 0.99 | 18.32 | 10.27 | 1857.47 | 15.77 | 1.67 | 6.96 | 6.96 | 1.14 | 7.51 | 4.68 |
| $Co_6Cr_{44}Fe_{23}Mn_{26}$ | -0.32 | 1.19 | 1.40 | 0.87 | 60.33 | 10.20 | 1893.62 | 14.53 | 1.68 | 6.61 | 6.92 | 0.96 | 7.52 | 7.90 |
| $Cr_{24}Fe_{30}Mn_{44}Ni_1$ | 0.26 | 1.01 | 1.40 | 0.45 | 65.57 | 9.51 | 1773.73 | 14.95 | 1.66 | 7.26 | 7.07 | 0.81 | 7.53 | 2.95 |
| $Al_{12}Cr_{18}Fe_{20}Mo_{37}Ti_{11}V_3$ | -7.98 | 4.34 | 1.40 | 4.28 | 3.65 | 13.22 | 2203.34 | 28.76 | 1.86 | 13.23 | 5.81 | 1.50 | 7.54 | 10.67 |
| $Cr_5Fe_{20}Mn_{38}Ni_{11}Sn_{26}$ | -1.93 | 3.94 | 1.41 | 2.09 | 8.37 | 11.85 | 1364.08 | 39.45 | 1.76 | 10.05 | 9.32 | 2.96 | 7.55 | 10.34 |
| $Al_9Cr_{42}Fe_{17}Mo_{26}Ti_3Zr_4$ | -6.46 | 4.18 | 1.40 | 4.19 | 4.07 | 12.06 | 2181.04 | 25.01 | 1.80 | 13.02 | 5.94 | 1.32 | 7.55 | 10.29 |
| $Cr_{20}Cu_5Fe_{35}Ni_{30}Ti_{10}$ | -8.86 | 6.87 | 1.38 | 1.76 | 2.50 | 11.96 | 1851.52 | 10.86 | 1.79 | 6.90 | 7.91 | 2.02 | 7.55 | 5.04 |
| $Cr_{24}Fe_{23}Mn_{44}Nb_6V_3$ | -1.52 | 2.33 | 1.40 | 1.10 | 13.37 | 11.02 | 1841.43 | 19.51 | 1.65 | 6.66 | 6.80 | 0.90 | 7.55 | 7.57 |

| Composition | | | | | | | | | | | | | |
|---|---|---|---|---|---|---|---|---|---|---|---|---|---|
| $Co_3Cr_{23}Fe_{39}Mn_{34}Ti_1$ | -0.60 | 1.60 | 1.40 | 0.60 | 30.40 | 10.16 | 1794.77 | 13.64 | 1.69 | 7.43 | 7.19 | 0.89 | 7.55 | 4.28 |
| $Cr_{28}Fe_{33}Mn_{28}Ni_7Zr_3$ | -3.51 | 3.99 | 1.40 | 2.11 | 5.98 | 11.44 | 1838.84 | 13.99 | 1.69 | 8.28 | 7.16 | 1.23 | 7.55 | 4.51 |
| $Cr_{42}Fe_{28}Mn_6Ni_{15}V_7$ | -4.32 | 2.44 | 1.39 | 1.49 | 5.18 | 11.42 | 1960.02 | 11.53 | 1.74 | 6.47 | 7.19 | 1.52 | 7.56 | 6.22 |
| $Cr_{24}Fe_{21}Mn_{44}Mo_5V_6$ | 0.42 | 1.33 | 1.40 | 1.20 | 50.09 | 11.27 | 1850.88 | 19.98 | 1.67 | 9.24 | 6.79 | 0.84 | 7.57 | 5.26 |
| $Co_2Cr_{23}Fe_{39}Mn_{36}$ | -0.06 | 1.07 | 1.40 | 0.70 | 280.43 | 9.64 | 1788.58 | 14.03 | 1.69 | 7.38 | 7.19 | 0.83 | 7.57 | 3.62 |
| $Co_{10}Cr_{33}Fe_{40}Mn_{13}Ti_2$ | -2.51 | 2.41 | 1.39 | 1.11 | 8.53 | 11.32 | 1891.99 | 11.90 | 1.73 | 6.66 | 7.23 | 1.17 | 7.57 | 9.16 |
| $Cr_{25}Cu_4Fe_{31}Mn_{40}$ | 1.78 | 1.87 | 1.40 | 0.68 | 9.89 | 9.96 | 1767.88 | 15.39 | 1.68 | 7.47 | 7.21 | 1.05 | 7.58 | 2.91 |
| $Cr_{24}Cu_5Fe_{26}Mn_{40}Mo_2V_4$ | 1.79 | 2.26 | 1.40 | 1.13 | 11.66 | 11.65 | 1796.16 | 17.81 | 1.68 | 8.24 | 7.12 | 1.19 | 7.58 | 4.16 |
| $Co_3Cr_{16}Fe_{28}Mn_{34}Ni_{13}Ti_4$ | -4.68 | 3.71 | 1.39 | 1.33 | 4.80 | 12.74 | 1765.45 | 12.76 | 1.71 | 8.41 | 7.47 | 1.38 | 7.59 | 5.51 |
| $Al_{14}Co_2Cr_{21}Fe_{32}Mn_3Mo_{24}Ni_5$ | -6.36 | 3.43 | 1.39 | 4.37 | 4.31 | 13.56 | 2017.94 | 30.50 | 1.84 | 11.09 | 6.50 | 1.78 | 7.59 | 9.12 |
| $Al_2Co_9Cr_4Cu_{10}Fe_{32}Mn_4Ni_{12}Ti_6$ | -3.23 | 5.92 | 1.38 | 2.08 | 8.39 | 14.88 | 1820.18 | 14.71 | 1.78 | 6.87 | 7.81 | 1.95 | 7.59 | 9.82 |
| $Cr_{40}Fe_{14}Mn_{21}Mo_{12}V_6Zr_7$ | -3.38 | 3.87 | 1.41 | 3.10 | 8.08 | 13.16 | 2072.87 | 19.27 | 1.69 | 12.30 | 6.28 | 1.03 | 7.60 | 9.64 |
| $Al_4Cr_{19}Cu_6Fe_{27}Mn_{16}Ni_{26}V_3$ | -4.85 | 4.61 | 1.38 | 2.62 | 5.17 | 14.26 | 1758.98 | 16.89 | 1.76 | 7.72 | 7.84 | 1.95 | 7.60 | 5.32 |
| $Cr_{23}Fe_{27}Mn_{43}Nb_3Ni_3$ | -1.22 | 2.26 | 1.40 | 0.88 | 15.56 | 10.57 | 1799.32 | 17.33 | 1.66 | 7.37 | 7.06 | 0.95 | 7.61 | 5.10 |
| $Co_9Cr_{32}Fe_{32}Mn_{26}V_1$ | -1.06 | 1.36 | 1.39 | 1.11 | 19.72 | 11.29 | 1852.72 | 13.79 | 1.71 | 7.13 | 7.16 | 1.02 | 7.61 | 9.11 |
| $Al_3Cr_{19}Cu_8Fe_{30}Ni_{32}Sn_2Ti_5$ | -5.54 | 6.43 | 1.38 | 2.56 | 4.22 | 13.22 | 1769.48 | 18.36 | 1.81 | 6.48 | 8.25 | 2.26 | 7.62 | 6.03 |
| $Cr_{36}Fe_{27}Mn_{30}Mo_2Ni_5$ | -0.50 | 1.60 | 1.40 | 0.95 | 40.83 | 10.91 | 1879.85 | 16.63 | 1.70 | 8.07 | 7.02 | 1.04 | 7.62 | 4.36 |
| $Co_{10}Cr_{28}Cu_{10}Fe_{27}Mn_{14}V_8Zr_1$ | 0.49 | 4.40 | 1.39 | 2.07 | 54.57 | 14.24 | 1858.78 | 15.47 | 1.73 | 7.45 | 7.41 | 1.70 | 7.63 | 11.59 |
| $Al_1Cr_{10}Cu_{13}Fe_{24}Mn_{27}Ni_{17}Ti_5$ | -2.81 | 5.72 | 1.38 | 1.96 | 8.90 | 14.75 | 1693.24 | 14.74 | 1.74 | 8.79 | 7.98 | 1.94 | 7.64 | 4.59 |
| $Co_4Cr_{28}Fe_{42}Mn_{20}Ni_4V_1$ | -1.36 | 1.42 | 1.40 | 1.03 | 15.30 | 11.22 | 1858.74 | 12.67 | 1.73 | 6.92 | 7.32 | 1.08 | 7.65 | 5.89 |
| $Cr_{24}Fe_{32}Mn_{21}Ni_{18}Ti_2V_3$ | -4.45 | 3.17 | 1.39 | 1.46 | 5.18 | 12.56 | 1837.95 | 12.83 | 1.73 | 7.70 | 7.52 | 1.47 | 7.66 | 4.64 |
| $Al_{10}Co_2Cr_{17}Cu_{16}Fe_{31}Mn_1Mo_9Ni_{14}$ | 0.00 | 5.81 | 1.37 | 3.67 | 6824.70 | 15.02 | 1796.14 | 27.29 | 1.83 | 8.16 | 7.76 | 2.33 | 7.66 | 7.13 |
| $Cr_{22}Cu_7Fe_{33}Mn_{31}Ni_7Ti_1$ | 0.52 | 3.36 | 1.39 | 1.23 | 42.15 | 12.36 | 1767.59 | 14.76 | 1.71 | 7.97 | 7.54 | 1.43 | 7.67 | 3.45 |
| $Cr_9Cu_{18}Fe_{21}Mn_{25}Ni_{18}Ti_7V_3$ | -2.69 | 6.07 | 1.38 | 1.77 | 9.34 | 14.83 | 1692.93 | 14.66 | 1.75 | 8.91 | 8.12 | 2.06 | 7.68 | 4.84 |
| $Al_3Cr_{16}Cu_{11}Fe_{36}Mn_9Ni_{21}Ti_3$ | -2.21 | 5.54 | 1.38 | 2.40 | 10.98 | 13.88 | 1749.41 | 15.82 | 1.79 | 7.03 | 8.09 | 1.97 | 7.68 | 4.36 |
| $Cr_{17}Cu_2Fe_{30}Mn_{19}Ni_{28}Si_3$ | -7.89 | 6.74 | 1.38 | 3.94 | 2.85 | 12.60 | 1783.09 | 11.87 | 1.77 | 7.74 | 7.97 | 1.64 | 7.68 | 4.64 |
| $Cr_{15}Cu_{19}Fe_{20}Mn_{20}Ni_{19}Ti_8$ | -2.20 | 6.32 | 1.38 | 1.75 | 11.36 | 14.56 | 1715.13 | 15.33 | 1.75 | 8.65 | 8.14 | 2.11 | 7.69 | 4.76 |
| $Co_3Cu_3Fe_{40}Mn_3Ni_{30}Ti_3V_{17}$ | -8.88 | 5.42 | 1.37 | 1.82 | 2.56 | 12.40 | 1835.01 | 10.29 | 1.80 | 6.46 | 8.02 | 1.90 | 7.69 | 8.78 |

| Composition | | | | | | | | | | | | | |
|---|---|---|---|---|---|---|---|---|---|---|---|---|---|
| $Al_4Co_4Cr_6Cu_{17}Fe_{17}Mn_{15}Ni_{31}Ti_5$ | -5.77 | 6.95 | 1.37 | 2.56 | 4.38 | 15.30 | 1651.62 | 15.70 | 1.79 | 8.25 | 8.47 | 2.22 | 7.70 | 7.61 |
| $Al_8Cr_{27}Fe_{33}Mn_{11}Mo_{21}$ | -3.33 | 3.02 | 1.40 | 3.49 | 7.59 | 12.46 | 2026.81 | 27.31 | 1.80 | 11.43 | 6.53 | 1.38 | 7.70 | 7.28 |
| $Cr_{10}Cu_5Fe_{35}Mn_9Ni_{33}Ti_9$ | -8.08 | 6.59 | 1.38 | 1.75 | 2.81 | 12.71 | 1785.96 | 10.28 | 1.79 | 7.47 | 8.16 | 1.88 | 7.71 | 4.68 |
| $Cr_{25}Fe_{28}Mn_{36}Mo_2Nb_8$ | -1.98 | 2.73 | 1.41 | 1.08 | 10.81 | 11.30 | 1897.92 | 20.74 | 1.67 | 8.03 | 6.84 | 0.92 | 7.71 | 8.44 |
| $Al_4Cr_{26}Fe_{23}Mn_{14}Mo_{20}V_{12}$ | -2.79 | 2.88 | 1.40 | 3.01 | 10.59 | 14.14 | 2091.74 | 24.22 | 1.78 | 11.92 | 6.35 | 1.20 | 7.71 | 9.55 |
| $Cr_4Fe_{14}Mn_{22}Mo_{34}Ti_{19}Zr_7$ | -5.68 | 4.53 | 1.43 | 2.84 | 5.10 | 13.31 | 2177.09 | 25.28 | 1.78 | 16.52 | 5.98 | 1.34 | 7.72 | 10.62 |
| $Cr_{27}Fe_{28}Ni_{32}V_{13}$ | -7.81 | 3.49 | 1.38 | 1.81 | 2.77 | 11.17 | 1933.66 | 10.64 | 1.78 | 6.38 | 7.70 | 1.86 | 7.72 | 7.88 |
| $Al_{10}Cr_{23}Fe_{28}Mo_{28}V_8$ | -5.31 | 2.77 | 1.39 | 4.10 | 5.15 | 12.72 | 2150.48 | 27.56 | 1.85 | 11.66 | 6.19 | 1.48 | 7.72 | 10.48 |
| $Cr_{13}Cu_9Fe_{35}Mn_5Ni_{28}Ti_8V_2$ | -5.83 | 6.70 | 1.38 | 1.78 | 4.16 | 13.52 | 1794.28 | 11.92 | 1.80 | 7.03 | 8.19 | 1.97 | 7.72 | 5.05 |
| $Cr_{28}Cu_5Fe_{24}Ni_{34}Ti_6V_2$ | -7.15 | 5.96 | 1.38 | 1.79 | 3.21 | 12.19 | 1878.91 | 11.94 | 1.79 | 6.82 | 7.98 | 2.02 | 7.72 | 6.35 |
| $Al_3Co_1Cr_{25}Fe_{32}Mn_2Ni_{35}Ti_1$ | -6.96 | 3.79 | 1.38 | 2.40 | 3.07 | 11.58 | 1842.10 | 13.53 | 1.80 | 6.37 | 7.97 | 1.86 | 7.73 | 6.21 |
| $Cr_{14}Cu_7Fe_{20}Mn_{41}Ni_{14}V_4$ | -1.03 | 3.43 | 1.39 | 1.57 | 21.53 | 12.97 | 1714.24 | 15.00 | 1.70 | 8.72 | 7.69 | 1.57 | 7.74 | 4.53 |
| $Co_9Cr_{11}Cu_{11}Fe_{22}Mn_{32}Ni_6V_9$ | 0.03 | 3.75 | 1.38 | 1.72 | 1018.49 | 14.78 | 1730.22 | 15.53 | 1.72 | 8.44 | 7.73 | 1.68 | 7.74 | 10.00 |
| $Al_2Co_5Cr_{28}Fe_{40}Mn_{15}Mo_5Ni_6$ | -2.46 | 2.23 | 1.39 | 1.98 | 9.87 | 12.79 | 1901.34 | 17.89 | 1.76 | 8.38 | 7.26 | 1.28 | 7.74 | 7.15 |
| $Co_7Cr_{28}Fe_{41}Mn_{18}Mo_4$ | -1.09 | 1.52 | 1.40 | 1.47 | 20.07 | 11.50 | 1902.15 | 16.81 | 1.74 | 8.23 | 7.18 | 1.04 | 7.74 | 7.71 |
| $Al_3Cr_4Cu_{23}Fe_{42}Mn_{12}Ni_{12}Ti_4$ | 1.88 | 6.04 | 1.38 | 2.42 | 11.49 | 13.11 | 1652.06 | 15.51 | 1.80 | 7.04 | 8.43 | 2.02 | 7.74 | 3.40 |
| $Al_3Cr_2Fe_{13}Mn_{37}V_{15}W_5$ | -1.59 | 2.86 | 1.39 | 2.22 | 15.34 | 12.64 | 1924.63 | 26.79 | 1.67 | 10.62 | 6.39 | 1.07 | 7.74 | 9.28 |
| $Al_2Cr_4Cu_{26}Fe_{13}Mn_{21}Mo_{10}Ni_1V_{24}$ | 4.40 | 4.82 | 1.38 | 2.70 | 6.05 | 14.65 | 1818.96 | 26.70 | 1.77 | 10.63 | 7.54 | 2.35 | 7.74 | 9.43 |
| $Cr_{13}Cu_8Fe_{21}Mn_{41}Nb_5Ni_9Zr_3$ | -3.29 | 5.40 | 1.40 | 2.47 | 7.24 | 13.64 | 1748.96 | 18.82 | 1.68 | 9.29 | 7.49 | 1.63 | 7.75 | 7.12 |
| $Cr_{28}Fe_{24}Mn_{43}Ni_3W_2$ | 0.19 | 1.42 | 1.40 | 0.73 | 98.16 | 10.22 | 1817.55 | 19.76 | 1.67 | 8.71 | 7.03 | 0.90 | 7.76 | 4.42 |
| $Cr_{39}Fe_{31}Mn_{14}Mo_{11}V_6$ | -0.80 | 1.34 | 1.40 | 1.46 | 30.35 | 11.77 | 2054.49 | 18.25 | 1.75 | 9.83 | 6.69 | 0.98 | 7.76 | 7.01 |
| $Al_2Cr_{21}Fe_{37}Mn_7Ni_{31}Ti_1$ | -6.12 | 3.60 | 1.38 | 2.21 | 3.46 | 11.62 | 1822.70 | 12.83 | 1.79 | 6.73 | 7.96 | 1.72 | 7.76 | 4.84 |
| $Cr_{16}Fe_{39}Mn_{32}Ni_{14}$ | -2.33 | 1.90 | 1.39 | 1.25 | 8.16 | 10.79 | 1763.97 | 12.28 | 1.73 | 8.08 | 7.66 | 1.18 | 7.77 | 3.17 |
| $Cr_{17}Cu_{17}Fe_{27}Mn_9Ni_{17}Ti_1V_{11}$ | -0.05 | 5.29 | 1.38 | 1.81 | 539.37 | 14.99 | 1799.83 | 16.00 | 1.78 | 7.24 | 8.04 | 2.05 | 7.78 | 6.79 |
| $Cr_{12}Fe_{40}Mn_{36}Nb_{12}$ | -3.96 | 3.35 | 1.41 | 1.15 | 4.85 | 10.31 | 1860.73 | 20.72 | 1.68 | 7.51 | 7.05 | 1.00 | 7.78 | 9.26 |
| $Cr_{19}Fe_{36}Mo_1Ni_{30}V_{14}$ | -7.44 | 3.43 | 1.38 | 1.87 | 2.93 | 11.37 | 1919.26 | 11.44 | 1.80 | 6.36 | 7.78 | 1.79 | 7.78 | 7.43 |
| $Al_9Cu_{38}Fe_{33}Ni_{20}$ | 3.93 | 5.74 | 1.36 | 3.04 | 4.17 | 10.63 | 1540.62 | 18.15 | 1.85 | 4.58 | 9.07 | 2.33 | 7.79 | 4.54 |
| $Co_6Cr_{11}Cu_{14}Fe_{32}Mn_{23}Ni_9Ti_3$ | 0.36 | 4.93 | 1.38 | 1.80 | 68.84 | 14.70 | 1708.14 | 14.31 | 1.76 | 8.13 | 8.06 | 1.75 | 7.79 | 7.24 |

| Composition | | | | | | | | | | | | | |
|---|---|---|---|---|---|---|---|---|---|---|---|---|---|
| $Al_3Cr_{13}Cu_{20}Fe_{28}Mn_9Ni_{21}Ti_2V_4$ | 0.21 | 5.72 | 1.37 | 2.31 | 118.72 | 14.81 | 1715.64 | 16.95 | 1.80 | 7.09 | 8.37 | 2.10 | 7.81 | 5.34 |
| $Co_1Cr_{26}Fe_{31}Mn_{13}Ni_{28}Si_1$ | -5.70 | 4.32 | 1.38 | 2.71 | 3.90 | 12.04 | 1842.65 | 11.87 | 1.77 | 7.25 | 7.87 | 1.56 | 7.81 | 5.79 |
| $Co_5Cr_{32}Fe_{39}Mn_{10}Mo_1Ni_{13}$ | -2.76 | 1.67 | 1.39 | 1.45 | 8.18 | 11.86 | 1901.02 | 12.60 | 1.76 | 6.87 | 7.55 | 1.34 | 7.81 | 7.48 |
| $Co_4Cr_{14}Cu_4Fe_{26}Mn_{26}Ni_{23}Ti_3$ | -4.23 | 4.34 | 1.38 | 1.66 | 5.73 | 13.85 | 1751.53 | 12.60 | 1.75 | 8.49 | 7.95 | 1.61 | 7.82 | 6.40 |
| $Cr_{19}Fe_{26}Mn_{10}Mo_{15}V_{28}W_1$ | -2.82 | 2.11 | 1.39 | 2.35 | 10.31 | 13.58 | 2143.38 | 20.31 | 1.77 | 11.51 | 6.36 | 1.18 | 7.82 | 11.57 |
| $Co_2Cr_{16}Fe_{39}Mn_7Ni_{31}Ti_4$ | -6.82 | 4.56 | 1.38 | 1.70 | 3.20 | 11.92 | 1828.60 | 9.54 | 1.80 | 6.76 | 8.08 | 1.63 | 7.83 | 5.89 |
| $Cr_{37}Fe_{13}Mn_{14}Mo_{19}Nb_1V_{16}$ | -0.91 | 1.70 | 1.40 | 2.13 | 31.31 | 13.03 | 2180.39 | 19.40 | 1.76 | 11.88 | 6.24 | 0.89 | 7.84 | 11.01 |
| $Co_2Cr_{11}Cu_5Fe_{28}Mn_{17}Ni_{35}Si_3$ | -6.97 | 6.91 | 1.37 | 3.94 | 3.29 | 13.13 | 1746.84 | 11.07 | 1.80 | 7.44 | 8.35 | 1.68 | 7.84 | 5.80 |
| $Al_5Co_4Cr_{14}Cu_{18}Fe_{28}Ni_{27}Zn_3$ | 0.37 | 5.22 | 1.37 | 2.70 | 62.75 | 13.92 | 1680.13 | 20.20 | 1.83 | 5.54 | 8.67 | 2.15 | 7.84 | 7.32 |
| $Cr_{31}Cu_7Fe_{30}Mn_{22}Mo_5Ni_4$ | 2.12 | 3.13 | 1.40 | 1.43 | 11.32 | 12.72 | 1883.19 | 19.24 | 1.74 | 8.99 | 7.34 | 1.43 | 7.84 | 4.87 |
| $Cr_{14}Fe_{34}Mn_{13}Ni_{31}V_8$ | -6.26 | 3.28 | 1.38 | 1.78 | 3.61 | 12.34 | 1829.90 | 11.53 | 1.78 | 7.29 | 7.97 | 1.64 | 7.86 | 6.12 |
| $Co_{13}Cr_{24}Fe_{44}Mn_2Ni_{15}Ti_2$ | -4.00 | 2.75 | 1.39 | 1.61 | 5.48 | 11.67 | 1879.12 | 9.53 | 1.80 | 5.65 | 7.85 | 1.40 | 7.87 | 11.41 |
| $Co_4Cr_4Cu_{23}Fe_{27}Ni_{23}Sn_5Ti_4V_{11}$ | -0.47 | 6.62 | 1.37 | 2.13 | 53.14 | 14.85 | 1681.88 | 21.78 | 1.83 | 6.01 | 8.93 | 2.37 | 7.87 | 9.99 |
| $Cr_{16}Fe_{32}Mn_{27}Ni_{24}Zn_1$ | -3.41 | 2.47 | 1.39 | 1.58 | 6.15 | 12.01 | 1749.77 | 14.23 | 1.74 | 8.20 | 7.97 | 1.46 | 7.87 | 4.06 |
| $Co_9Cr_{24}Cu_3Fe_{18}Ni_{44}Si_4$ | -8.63 | 6.87 | 1.36 | 4.15 | 2.55 | 11.92 | 1844.39 | 10.97 | 1.83 | 5.54 | 8.40 | 1.86 | 7.88 | 11.37 |
| $Al_5Cr_8Cu_4Fe_{23}Mn_{42}Mo_{18}$ | 0.56 | 4.11 | 1.40 | 2.90 | 40.64 | 12.41 | 1844.55 | 29.68 | 1.75 | 12.87 | 6.95 | 1.40 | 7.88 | 6.05 |
| $Cr_{22}Fe_{21}Mn_{21}Mo_{14}Nb_6Ni_2V_{13}$ | -3.22 | 3.12 | 1.40 | 2.14 | 9.70 | 14.96 | 2088.33 | 21.93 | 1.74 | 11.32 | 6.53 | 1.17 | 7.89 | 11.96 |
| $Al_8Cr_{28}Fe_{27}Mn_4Mo_{27}Nb_3V_2$ | -4.59 | 3.04 | 1.40 | 3.68 | 6.17 | 13.14 | 2158.19 | 26.78 | 1.83 | 11.90 | 6.29 | 1.36 | 7.89 | 10.46 |
| $Cr_{32}Fe_{26}Mn_{11}Ni_{30}$ | -4.51 | 2.16 | 1.38 | 1.66 | 4.56 | 10.99 | 1872.06 | 12.11 | 1.77 | 7.18 | 7.86 | 1.61 | 7.90 | 5.51 |
| $Co_3Cu_{10}Fe_{35}Mn_{19}Ni_{26}Ti_5V_2$ | -3.89 | 5.75 | 1.38 | 1.79 | 5.81 | 13.27 | 1700.24 | 10.28 | 1.79 | 7.94 | 8.42 | 1.72 | 7.90 | 5.98 |
| $Co_9Cu_{17}Fe_{24}Mn_{22}Ni_{15}V_{12}$ | -1.13 | 4.90 | 1.37 | 1.82 | 22.06 | 14.73 | 1698.03 | 14.31 | 1.77 | 8.21 | 8.29 | 1.89 | 7.90 | 10.86 |
| $Co_9Cr_{41}Fe_{17}Ni_{30}V_3$ | -5.75 | 2.33 | 1.38 | 1.79 | 3.77 | 11.16 | 1942.73 | 10.82 | 1.78 | 6.37 | 7.80 | 1.77 | 7.91 | 12.00 |
| $Co_3Cr_{27}Cu_2Fe_{26}Mn_7Ni_{30}V_4$ | -4.65 | 3.25 | 1.38 | 1.77 | 5.27 | 13.09 | 1871.75 | 12.16 | 1.78 | 6.84 | 7.95 | 1.73 | 7.91 | 7.96 |
| $Co_2Cr_{21}Fe_{33}Mn_{16}Ni_{26}Zn_2$ | -3.79 | 2.22 | 1.39 | 1.65 | 5.87 | 12.34 | 1802.32 | 14.23 | 1.77 | 7.38 | 8.02 | 1.53 | 7.91 | 5.64 |
| $Co_4Cr_{21}Cu_{12}Fe_{28}Mn_6Ni_{26}Ti_4$ | -2.03 | 5.67 | 1.38 | 1.79 | 12.57 | 14.21 | 1796.25 | 13.73 | 1.80 | 6.89 | 8.29 | 1.90 | 7.91 | 7.31 |
| $Cr_{41}Fe_{20}Mo_{25}Ti_6V_7$ | -3.20 | 2.16 | 1.41 | 1.92 | 8.25 | 11.61 | 2272.92 | 17.15 | 1.81 | 11.99 | 6.21 | 1.04 | 7.91 | 10.20 |
| $Cr_{25}Fe_{28}Mn_9Ni_{37}Si_1$ | -6.13 | 4.11 | 1.38 | 2.71 | 3.33 | 11.07 | 1847.01 | 11.24 | 1.79 | 6.86 | 8.10 | 1.65 | 7.92 | 5.58 |
| $Co_2Cr_{11}Cu_{17}Fe_{31}Mn_4Ni_{21}V_{13}$ | -0.20 | 5.17 | 1.37 | 1.82 | 128.27 | 14.29 | 1794.83 | 14.99 | 1.80 | 6.31 | 8.30 | 1.99 | 7.92 | 8.18 |

| Composition | | | | | | | | | | | | | |
|---|---|---|---|---|---|---|---|---|---|---|---|---|---|
| $Cr_{21}Cu_{30}Fe_{18}Ni_{22}Ti_6V_2$ | 1.19 | 6.72 | 1.37 | 1.81 | 19.86 | 13.47 | 1750.82 | 17.25 | 1.81 | 6.63 | 8.62 | 2.27 | 7.92 | 6.63 |
| $Co_2Cr_{18}Fe_{24}Mn_{20}Ni_{33}V_3$ | -5.57 | 2.92 | 1.38 | 1.76 | 4.13 | 12.77 | 1803.52 | 12.13 | 1.77 | 8.00 | 8.03 | 1.60 | 7.93 | 6.87 |
| $Cr_8Cu_{10}Fe_{38}Mn_{15}Ni_{22}V_7$ | -0.95 | 4.26 | 1.38 | 1.77 | 24.76 | 13.36 | 1755.89 | 13.12 | 1.78 | 7.28 | 8.23 | 1.69 | 7.93 | 5.08 |
| $Al_7Cr_{14}Cu_{19}Fe_{28}Mo_8Ni_{23}$ | 0.84 | 5.77 | 1.37 | 3.30 | 29.48 | 13.90 | 1780.23 | 25.34 | 1.85 | 7.23 | 8.26 | 2.27 | 7.94 | 6.41 |
| $Al_1Cr_{15}Cu_4Fe_{33}Ni_{44}Si_2$ | -5.92 | 6.09 | 1.37 | 3.69 | 3.25 | 10.69 | 1797.43 | 11.19 | 1.84 | 4.80 | 8.57 | 1.76 | 7.94 | 5.69 |
| $Co_3Cr_6Cu_{23}Fe_{25}Mn_{13}Ni_{23}Ti_6$ | -1.03 | 6.55 | 1.38 | 1.82 | 23.48 | 14.47 | 1678.33 | 13.63 | 1.80 | 7.69 | 8.68 | 2.00 | 7.94 | 6.35 |
| $Al_3Cr_{24}Fe_{27}Mn_8Mo_{24}Ni_2V_{12}$ | -3.18 | 2.62 | 1.40 | 2.88 | 9.44 | 13.95 | 2149.07 | 23.01 | 1.81 | 11.74 | 6.48 | 1.27 | 7.95 | 9.85 |
| $Co_8Cu_{14}Fe_{34}Mn_{25}Ni_{18}Si_2$ | -2.08 | 6.56 | 1.37 | 3.51 | 10.41 | 13.10 | 1655.69 | 9.88 | 1.79 | 7.89 | 8.51 | 1.54 | 7.95 | 7.86 |
| $Co_7Cu_4Fe_{39}Mn_{21}Ni_{26}Ti_3$ | -4.55 | 4.66 | 1.38 | 1.75 | 4.68 | 12.43 | 1715.78 | 8.23 | 1.79 | 7.80 | 8.34 | 1.46 | 7.95 | 7.47 |
| $Cr_{21}Fe_{31}Mn_{26}Mo_8Ni_{13}V_1$ | -2.26 | 2.23 | 1.40 | 1.65 | 10.78 | 12.89 | 1891.86 | 19.58 | 1.76 | 9.91 | 7.39 | 1.28 | 7.96 | 5.40 |
| $Al_5Co_8Cu_{35}Fe_{19}Ni_{23}V_{11}$ | 0.81 | 5.85 | 1.35 | 2.25 | 26.99 | 13.41 | 1630.55 | 18.58 | 1.84 | 5.36 | 9.01 | 2.29 | 7.96 | 11.55 |
| $Co_7Cr_8Cu_{36}Fe_{28}Ni_{11}Ti_7V_2$ | 2.63 | 6.84 | 1.37 | 1.81 | 8.67 | 13.58 | 1681.81 | 16.14 | 1.83 | 5.98 | 8.87 | 2.17 | 7.96 | 8.83 |
| $Al_5Cr_{25}Fe_{17}Mo_{26}Ni_{12}V_{14}$ | -7.10 | 3.77 | 1.39 | 3.54 | 4.40 | 14.29 | 2188.91 | 23.47 | 1.84 | 11.47 | 6.53 | 1.72 | 7.96 | 11.87 |
| $Cu_{16}Fe_{22}Mn_{14}Ni_{34}Sn_1V_{11}$ | -3.59 | 6.18 | 1.37 | 2.69 | 6.35 | 13.43 | 1695.92 | 15.45 | 1.81 | 7.48 | 8.70 | 2.01 | 7.97 | 7.24 |
| $Al_2Co_7Cr_{17}Fe_{35}Mn_3Ni_{35}$ | -4.86 | 2.82 | 1.38 | 2.28 | 4.51 | 12.10 | 1809.56 | 12.12 | 1.82 | 5.79 | 8.28 | 1.65 | 7.97 | 9.30 |
| $Cr_{17}Fe_{18}Mn_{31}Mo_{20}V_{13}$ | 0.26 | 1.81 | 1.40 | 2.04 | 104.79 | 13.05 | 2049.30 | 24.30 | 1.75 | 12.85 | 6.55 | 0.93 | 7.97 | 9.07 |
| $Cr_7Cu_{24}Fe_9Mn_{37}Ni_{18}Ti_1Zr_3$ | -1.50 | 6.01 | 1.38 | 2.69 | 14.10 | 13.12 | 1614.54 | 14.71 | 1.73 | 10.19 | 8.42 | 1.99 | 7.97 | 5.47 |
| $Co_6Cr_{28}Fe_{23}Ni_{40}Si_2$ | -6.90 | 4.56 | 1.37 | 3.06 | 2.94 | 10.81 | 1880.55 | 10.36 | 1.82 | 5.79 | 8.22 | 1.74 | 7.97 | 9.97 |
| $Cr_{16}Cu_{28}Fe_{18}Ni_{26}Si_1V_{10}$ | 0.64 | 6.72 | 1.36 | 2.86 | 36.58 | 13.31 | 1756.66 | 17.46 | 1.82 | 5.98 | 8.69 | 2.22 | 7.97 | 7.51 |
| $Cr_{19}Fe_{38}Mn_8Nb_3Ni_{28}V_2$ | -6.31 | 3.74 | 1.39 | 1.87 | 3.56 | 11.98 | 1873.86 | 13.33 | 1.79 | 6.69 | 7.96 | 1.59 | 7.98 | 6.99 |
| $Al_4Cr_{11}Cu_{34}Fe_{30}Ni_{19}V_3$ | 4.92 | 5.46 | 1.37 | 2.50 | 4.23 | 12.58 | 1654.67 | 18.86 | 1.84 | 5.19 | 8.89 | 2.18 | 7.98 | 5.36 |
| $Co_6Cr_{15}Fe_{28}Ni_{38}V_{11}$ | -7.40 | 3.59 | 1.37 | 1.85 | 3.07 | 12.09 | 1877.80 | 10.60 | 1.82 | 5.95 | 8.19 | 1.78 | 7.99 | 11.10 |